\documentstyle[preprint,aps,epsf]{revtex}

\newcommand{\be}{\begin{eqnarray}}
\newcommand{\ee}{\end{eqnarray}}

\newcommand{\dslash}{\partial\!\!\!/\,}
\newcommand{\Dslash}{D\!\!\!\!/\,}
\newcommand{\ts}{\textstyle}

\begin{document}
\bibliographystyle{unsrt}
\draft
\title{Instantons and the Chiral Phase Transition\\ 
       at non-zero Baryon Density}

\author{T.~Sch\"afer}

\address{Institute for Nuclear Theory, Department of Physics,
	 University of Washington,\\ Seattle, WA 98195, USA}

\maketitle

\begin{abstract}
  We study an interacting ensemble of instantons at finite
baryon chemical potential. We emphasize the importance of 
fermionic zero modes and calculate the fermion induced 
interaction between instantons at non-zero chemical potential. 
We show that unquenched simulations of the instanton ensemble
are feasible in two regimes, for sufficiently small and for
very large chemical potential. At very large chemical potential
chiral symmetry is restored and the instanton ensemble is 
dominated by strongly correlated chain-like configurations. 
 
\end{abstract}

\section{Introduction}
\label{sec_intro}

  The behavior of matter under extreme conditions of temperature
and/or density is of great interest in connection with the physics 
of heavy ion collisions and neutron stars. In addition to that,
understanding the nature of the phase transition between a hadronic
gas at low temperature and density and the quark gluon plasma at
high $T$ and $\rho_B$ is an important problem in its own right 
and expected to shed some light on the vacuum structure of QCD. 

  While the structure of QCD at high temperature has been investigated
in some detail using lattice simulations \cite{deT_95}, little is 
known about matter at high baryon density. This is unfortunate, 
because the possible phase structure of matter at finite density is 
much richer, including phenomena such as pion \cite{Mig_72} and kaon 
condensation \cite{KN_86}, density isomers \cite{Lee_75}, strange quark 
matter \cite{Wit_84}, etc. From the point of view of lattice QCD, the 
difficulty is connected with the sign problem (see \cite{BMK_97} for
a recent review on lattice simulations at finite chemical potential). 
At zero temperature and chemical potential, the weight factor 
$\det(\Dslash+m)\exp(-S)$ in the euclidean path integral is real 
and positive. This means that the weight function can be interpreted 
as a probability distribution and one can perform simulations based
on importance sampling. This remains true also at finite temperature. 
In euclidean space, the heat bath is represented by the (anti)periodic 
boundary conditions imposed on the gauge fields and fermions. At finite 
density, however, the determinant $\det(\Dslash+m-\mu\gamma_4)$ is 
complex and cannot be interpreted as a probability distribution. 

   Furthermore, while it is sometimes
argued that neglecting the determinant (the so-called quenched 
approximation) provides a reasonable approximation at finite
temperature, the same cannot be true at finite chemical potential. 
If the determinant is neglected, or if only the modulus of the 
determinant is retained, the critical chemical potential for 
chiral symmetry restoration is zero. This phenomenon is easily 
understood in the case where we retain the modulus only \cite{Goc_88b}.
This approximation corresponds to a theory where half of the 
quarks couple with the opposite chemical potential. In this 
case, the lightest state that carries baryon number is a
baryonic pion, a pion made of an ordinary quark and an anti-quark 
that couples with the opposite chemical potential. This state 
is degenerate with the pion, so that in the chiral limit it can
Bose condense even if the chemical potentials is arbitrarily small. 
In full QCD, on the other hand, the baryon density has to be zero 
as long as $\mu<\mu_c\simeq M_B/3$, where $M_B$ is the mass of the 
lightest baryon. 
 
 Recently, this phenomenon has been studied in some detail in
random matrix models of QCD \cite{Ste_96,JNP_96,HJV_97}. While these 
models nicely illustrate the shortcomings of the quenched 
approximation they do not really improve our understanding of the 
phase transition in full QCD. The chemical potential is simply added
as an external field to the random matrix that represents the
Dirac operator. Chiral symmetry is restored when the chemical
potential exceeds a certain critical value, but the dynamics
of the phase transition is equivalent to a simple mean field
model with static quarks \cite{JNZ_97}.

   In this work, we want to study QCD at finite chemical 
potential in the instanton liquid model. The instanton model
is based on the assumption that the physics of chiral symmetry
breaking and the dynamics of light hadrons is dominated by
just a few low-lying modes in the spectrum of the Dirac operator.
These modes are linear combinations of the zero modes associated
with instantons, classical solutions of the euclidean Yang-Mills
field equations. The instanton model provides a phenomenologically
very successful description of the QCD vacuum, chiral symmetry 
breaking and hadronic correlation functions \cite{SS_97}. In 
addition to that, the underlying assumptions of the model can be 
checked in great detail on the lattice, see e.g. \cite{CGHN_94,MS_95}. 
Of particular interest are recent simulations that focus on the 
correlation of low-lying eigenstates of the Dirac operator with 
topological charges, and on the role that these states play in the 
correlation functions of light hadrons \cite{TFMS_96,IN_97,DHK_97,VK_97}.
There is evidence that low-lying states are approximately linear 
combinations of instanton zero modes, and that these states play
a dominant role in the correlation functions of light mesons.

   Studying instantons at finite chemical potential is worth while
for several reasons. First, the model provides a very interesting
mechanism for the chiral phase transition at finite temperature,
based on the formation of topologically neutral instanton-anti-instanton
pairs near $T=T_c$ \cite{SSV_95}. The corresponding eigenstates are 
localized modes, contrary to the delocalized states that make up 
the quark condensate at $T=0$. This scenario raises the question 
whether a similar mechanism might be at work at finite baryon density. 
Second, because the model concentrates on just a few modes in 
the spectrum of the Dirac operator, the sign problem should 
not be as severe as it is in lattice simulations. This should
allow more realistic simulations to be performed. Finally, the 
instanton model can be studied in a number of approximations
that allow for analytical (or almost analytical) solutions. 
In this case, one might be able to handle the analytic continuation 
required to introduce a chemical potential.

   This paper is organized as follows. In Sect. 2 we discuss zero
modes, the anomaly and the density of instantons at $\mu\neq 0$.
In Sect. 3 we calculate matrix elements of the Dirac operator 
between zero mode wave functions. In Sect. 4, we study the 
spectrum of the Dirac operator in an ensemble of instantons 
at $\mu\neq 0$. In Sect. 5, we present simulations of the 
instanton liquid at small and large chemical potential.

\section{Zero modes and the anomaly at $\mu\neq 0$} 
\label{sec_zm}

  We would like to begin our discussion by constructing the free 
euclidean coordinate space propagators of bosons and fermions at 
non-zero chemical potential. This means that we are looking for 
solutions of the equations $(i\partial_\alpha-i\delta_{\alpha 4}
\mu)^2\Delta(x)=\delta(x)$ and $(i\dslash-i\mu\gamma_4)S(x)=\delta(x)$ 
subject to the appropriate boundary conditions. Naively, we can 
obtain solutions by multiplying the zero density propagator by 
$\exp(\mu x_4)$. This means that at finite chemical potential
the propagation of quarks along the positive time direction (and
anti-quarks along the negative time direction) is favored over
propagation in the opposite direction. 

  However, this naive prescription does not provide the correct 
propagator. We have missed the contribution from occupied states
below the Fermi surface. The correct solution is most easily obtained 
by Fourier transforming the momentum space propagator. The result 
for the charged Klein-Gordon propagator is
\be
\label{kg_mu}
\Delta(x) &=& \frac{1}{4\pi^2}\frac{1}{r^2+x_4^2}
 \left( \cos(\mu r) + \frac{x_4}{r} \sin(\mu r)\right).
\ee
Note that the extra contribution from occupied states has cancelled
the exponent $\exp(\mu x_4)$. From (\ref{kg_mu}) we can also obtain 
the Fermion propagator
\be 
\label{s_mu}
 S(x) &=& \left(i\dslash+i\mu\gamma_4\right)\Delta(x)
     \;=\; \frac{1}{4\pi^2x^4}\left(\vec\gamma\cdot\hat r S_r(x)
	   + \gamma_4 S_4(x) \right).
\ee
where the two functions $S_r(x)$ and $S_4(x)$ are given by
\be 
 S_r(x) &=& \left( 2r-\mu x^2x_4/r \right)\cos(\mu r)
    +\left( x_4^4+3r^2x_4 +\mu r^2 x^2 \right)\frac{\sin(\mu r)}{r^2},\\ 
 S_4(x) &=& \left( 2x_4+\mu x^2 \right)\cos(\mu r)
    +\left( x_4^2-r^2 +\mu x_4 x^2 \right)\frac{\sin(\mu r)}{r} .
\ee
Asymptotically, the propagator behaves as $S(x_4)\sim\mu^2/x_4$ in 
the temporal direction, as compared to $\sim 1/x_4^3$ at $\mu=0$. The 
propagator in the positive $x_4$ direction is still enhanced, but only 
by a power of $x_4$ rather than an exponential. This enhancement is 
related to the fact that at finite $\mu$, quark paths dominantly
wind around the imaginary time direction. The number of loops piercing
an $x_4=const$ surface going forward in time, minus the number of 
loops going backwards, is related to the total baryon number. From
(\ref{s_mu}) one easily reproduces the perturbative result $\rho_B
=4S_4(r\to 0,x_4=0)=\mu^3/(3\pi^2)$. In the next section we will 
also see that the anisotropy of the propagator causes the interaction
between instantons to be strongly anisotropic as well. 
 
  For spacelike separations the propagator behaves as $S(r)\sim\mu
\sin(\mu r)/r^2$. It decays as $1/r^2$ and oscillates with a wave 
number given by the Fermi momentum $p_F=\mu$. The oscillations are
due to the presence of a sharp Fermi surface and are analogous to
Friedel oscillations in a degenerate electron gas \cite{KT_88,FB_96}.

   Zero modes in the spectrum of the Dirac operator play a central 
role in understanding the $U(1)_A$ anomaly and the mechanism for
chiral symmetry breaking. While the relation to the anomaly is a 
rigorous consequence of the arguments given below, the connection
with chiral symmetry breaking is more subtle. The Banks-Casher 
formula \cite{BC_80} $\langle\bar qq\rangle =-\pi\rho(\lambda=0)$ 
relates the density of eigenvalues of the Dirac operator at zero 
virtuality to the quark condensate. In a finite system with volume 
$V$ these states are not exact zero modes, instead they occupy a 
band around zero with a typical level spacing $\Delta \lambda 
\sim |\langle\bar qq\rangle|/V$ \cite{LS_92}. Perturbative states, 
on the other hand, are spaced much farther apart, $\Delta \lambda 
\sim 1/V^{1/4}$. It is tempting to identify the extra states with 
quasi zero modes that arise when the exact zero modes of isolated 
instantons and anti-instantons interact. The fact that this mechanism 
leads to chiral symmetry breaking was checked in simulations of the 
instanton ensemble in QCD \cite{SV_90,Ver_94b}. Whether quasi zero 
modes associated with topology dominate the quark condensate in QCD is 
now under active investigation on the lattice \cite{IN_97,DHK_97,TFM_96}, 
and initial results are very promising. 

  The connection of zero modes with the axial anomaly can be seen 
from the fact that the violation of axial charge in an arbitrary
gauge potential is given by $\Delta Q_5=2N_f(n_L-n_R)$, where 
$n_{L,R}$ is the number of left and right handed zero modes of 
the Dirac operator. Index theorems (or the anomaly equation 
$\partial_\mu j_\mu^5=N_f/(16\pi^2)G_{\mu\nu} \tilde G_{\mu\nu}$) 
then imply that zero modes have to be associated with topological 
charges. This requirement is satisfied by the instanton solution. 
The instanton has topological charge $Q=+1$, and the Dirac operator 
in the instanton field has a left-handed zero mode, $i\Dslash\psi_0
=0$ and $\gamma_5\psi_0=-\psi_0$. Analogously, there is a right 
handed zero mode in the field of an anti-instanton.

   It is essential that these states are exact zero modes of the 
Dirac operator, because every non-zero mode automatically occurs
in pairs and only unpaired states contribute to $\Delta Q_5$.
If $\psi_\lambda$ is an eigenstate of $i\Dslash$ with eigenvalue 
$\lambda\neq 0$, then $\gamma_5\psi_\lambda$ is an eigenstate with 
eigenvalue $-\lambda$. This remains true at $\mu\neq 0$, because
the extra term $i\mu\gamma_4$ has the same chirality as $i\Dslash$.
But the anomaly is a short distance effect and should not be affected
by a small chemical potential\footnote{See \cite{Sch_96,LH_96,EHS_96} 
for a similar discussion at finite temperature.}. We therefore 
expect that there should be an exact zero mode in the spectrum of 
the Dirac operator in the instanton field also at $\mu\neq 0$. 

   This state can be found using the following observation 
\cite{Car_80}: Naively, a solution of the Dirac equation at $\mu
\neq 0$ is given by $\exp(\mu x_4)\psi_0(x)$, where $\psi_0(x)$
is the $\mu=0$ zero mode. From our discussion of the quark 
propagator we expect that this is not an acceptable solution. 
Instead, we have to construct a solution of the $\mu=0$
equation that cancels the exponential growth. 

   Such a solution can be found from the general ansatz \cite{Gro_77}
\be 
\label{zm_ans}
 \psi_0(x) &=& \frac{1}{2\sqrt{2}\pi\rho} \sqrt{\Pi(x)}
  \dslash \left( \frac{\Phi(x)}{\Pi(x)} \right)U^{ab} 
  \gamma_\pm \chi^b, 
\ee
where $\chi_i^a=\epsilon_{ia}$ is a Dirac and color spinor and 
$\gamma_\pm=(1\pm\gamma_5)/2$ for instantons/anti-instantons. The
gauge potential of the instanton is given by $A_\nu=\overline
\eta^{a}_{\nu\rho}\partial_\rho \ln\Pi(x)$ where $\overline 
\eta^{a}_{\mu\nu}$ is the anti-self-dual 't Hooft symbol 
\cite{tHo_76b} and $\Pi(x)=1+\rho^2/x^2$. For an anti-instanton, 
we have to make the usual replacement $\overline\eta^a_{\mu\nu}
\to\eta^a_{\mu\nu}$. Substituting the ansatz (\ref{zm_ans}) into 
the Dirac equation leads to the condition that $\Phi(x)$ has to 
be a solution of the Laplace equation $\Box\Phi(x)=0$. At $\mu=0$, 
a solution that is regular and satisfies the normalization
condition is given by $\Phi(x)=\rho^2/x^2$. 

  This suggests that a solution which satisfies the boundary condition
at finite chemical potential can be found from the $\mu\neq 0$ 
Klein-Gordon propagator $\Delta(x)$. If $\Delta(x)$ satisfies the 
Klein-Gordon equation with a chemical potential, then $\Phi(x)=
(4\pi^2)\rho^2\exp(-\mu x_4)\Delta(x)$ is a solution of the Laplace
equation that behaves as $\Phi(x)\sim\exp(-\mu x_4)$. It is easy to 
check that 
\be 
\label{zm_sol}
 \psi_0(x) &=& \sqrt{2}\pi\rho  \exp(\mu x_4)\sqrt{\Pi(x)}
  \dslash \left( \frac{\exp(-\mu x_4)\Delta(x)}{\Pi(x)} \right)U^{ab} 
  \gamma_\pm \chi^b
\ee
indeed solves $(i\Dslash-i\mu\gamma_4)\psi_0=0$ \cite{Car_80,Abr_83}. 
Furthermore, the zero mode solution is well behaved at infinity. It 
decays like the free quark propagator (\ref{s_mu}). Some care has to 
be taken in constructing the adjoint solution. The Dirac operator is 
not hermitean, so left eigenstates are not the hermitean conjugate 
of right eigenstates. In fact, it is easy to see that $\psi_0^\dagger
(\mu,x)=[\psi_0(-\mu,x)]^*$ is a solution of $\psi_0^\dagger(i\Dslash
-i\mu\gamma_4)=0$. Using this property, we can also check that the 
zero mode is correctly normalized,
\be
\int d^4x\, \psi_0^\dagger(\mu,x)\psi_0(\mu,x) &=& 1.
\ee

  In order to determine the importance of instanton effects at finite 
baryon density, we have to know the dependence of the instanton rate 
on the baryon chemical potential. At large chemical potential, this
dependence can be determined from a perturbative calculation. The 
result is\footnote{See \cite{Shu_82c} for a discussion of the  
disagreements among some of the earlier calculations.} 
\cite{Car_80,Bal_81,Che_81,Shu_82c,Abr_83}. 
\be
\label{dens_mu}
 n(\rho,\mu) &=& n(\rho,0)\exp\left(-N_f(\rho\mu)^2\right),
\ee
where $n(\rho,0)$ is the $\mu,T=0$ rate originally calculated by 
't Hooft \cite{tHo_76b}. The suppression factor $\exp(-N_f(\rho\mu)^2)$ 
is completely analogous to the factor $\exp(-2(N_c/3+N_f/6)(\pi\rho T)^2)$
that appears in the rate at finite $T$ \cite{GPY_81}. In both cases, the 
physical reason for the suppression of instantons is Debye screening of 
the strong $O(1/g)$ color electric fields inside the instanton. The rate
at $T\neq 0$ contains some extra terms that are due to the fact that
the instanton solution itself is deformed at $T\neq 0$. At least in
perturbation theory, this effect is absent at $\mu\neq 0$.

  Since the high temperature suppression factor is due to the scattering
of perturbative quarks and gluons on the instanton, it was argued that 
the result should not be trusted at temperatures around and below $T_c$
\cite{SV_94}. At low temperature the rate is determined by the scattering
of hadrons, not quarks and gluons, on the instanton. This has been 
checked in lattice calculations of the temperature dependence of the 
topological susceptibility in quenched QCD \cite{CS_95,IMM_95,ADD_96}.
These calculations find the expected exponential suppression above 
$T_c$, but below $T_c$ the topological susceptibility is essentially
independent of $T$.

  Analogously, we expect that the result (\ref{dens_mu}) is only valid
if the chemical potential is large. At small baryon density, the rate 
is determined by the density dependence of the effective $2N_f$-fermion 
operator induced by instantons. In the case of small $T$ and zero $\mu$
this dependence can be determined from soft pion theorems \cite{SV_94}.
At finite density, only the behavior of the quark condensate is known
\cite{DL_89}
\be
\langle \bar qq\rangle &=& \langle\bar qq\rangle \left\{
 1 - \frac{\Sigma_{\pi N}}{f_\pi^2 m_\pi^2}\rho_B + \ldots \right\},
\ee
where $\Sigma_{\pi N}\simeq 46$ MeV is the $\pi N$ sigma term. This 
implies $\langle\bar qq\rangle \simeq\langle\bar qq\rangle_0(1-0.3
(\rho_B/\rho_B^0))$ where $\rho_B^0=0.14\,{\rm fm}^{-3}$ is nuclear
matter density. If we estimate the density dependence of the 't Hooft 
operator using factorization, we get a very rapid dependence, $\langle 
{\cal O}_{'tH}\rangle \simeq\langle {\cal O}_{'tH}\rangle_0 (1-0.6
(\rho_B/\rho_B^0))$ for $N_f=2$. This seems too large, and factorization
is not expected to be a good approximation at finite density. In order
to improve on this estimate we have to determine the expectation 
value of the 't Hooft operator inside the nucleon. We have not 
performed this calculation, but it appears straightforward in principle. 
In practice, we have not used the perturbative result (\ref{dens_mu})
in our simulations at small chemical potential, but we have taken
it into account for large $\mu$. 

\section{Matrix elements of the Dirac operator}
\label{sec_tia}

  In order to study an ensemble of instantons at finite chemical 
potential, we have to determine matrix elements of the Dirac operator 
$(i\Dslash-i\mu\gamma_4)$ between the zero modes wave functions 
$\psi_{I}$ associated with different instantons. At finite temperature, 
the functional form of these matrix elements plays a crucial role 
in determining the mechanism for the chiral phase transition. 
Near $T_c$, the overlap matrix elements favor the formation of
polarized instanton-anti-instanton molecules \cite{IS_94,SSV_95}.
Our aim here is to identify instanton configurations that 
may play a similar role at large chemical potential. 

  In the following, we will assume that the gauge potential of a 
system of instantons and anti-instantons can be approximated by the 
sum of the individual gauge potentials $A_\nu=\sum_I A_\nu^I$. In
this case, the overlap matrix element is given by
\be
\label{def_me}
T_{IA} &=& \int d^4x\, 
  \psi_I^\dagger(x-z)\left( i\Dslash -i\mu\gamma_4\right) \psi_A(x) 
 \;=\;  \int d^4x\, 
  \psi_I^\dagger(x-z)\left( -i\dslash +i\mu\gamma_4\right) \psi_A(x).
\ee
Here, we have used the equations of motion to reduce the covariant
derivative to an ordinary one. Also note that because of the chirality 
of the zero modes, only matrix elements between instantons and 
anti-instantons are non-vanishing. 

  The overlap matrix elements $T_{IA}$ depend on the distance $z_\nu$ 
between the instantons, the instanton sizes $\rho_{I,A}$ and the 
relative color orientation $U=U_I^\dagger U_A$. In the following,
we will characterize the color orientation using the four-vector $u_\nu
=1/(2i){\rm tr}(U\tau_\nu^{(+)})$, where $\tau_\nu^{(\pm)}=(\vec\tau,
\mp)$. At $\mu,T=0$, Lorentz invariance implies that $T_{IA}=i(u\cdot
\hat z)f(|z|,\rho_I,\rho_A)$. At $\mu,T\neq 0$, Lorentz invariance is
broken and the matrix elements have the more general structure (see
\cite{SV_91,KY_91} for a discussion of $T_{IA}$ at $T\neq 0$) 
\be
 T_{IA} &=& iu_4 f_1 + i(\vec u \cdot \hat z) f_2
\ee
with $f_i=f_i(|\vec z|,z_4,\rho_I,\rho_A)$ and $\hat z=\vec z/|\vec z|$.
These functions can be calculated by writing the zero mode solutions in 
the form $\psi_i^a=\phi_\nu(x)(\gamma_\nu)_{ij}U^{ab}\chi^b_j$. At $\mu,
T=0$, we have $\phi_\nu(x)=\hat x_\nu \phi(|x|)$, while in the more 
general case $\phi_\nu(x)=\delta_{\nu 4}\phi_4(r,x_4)+\delta_{\nu i}
\hat r_i\phi_r(r,x_4)$. In terms of $\phi_\nu$, the overlap matrix 
elements are given by
\be
\label{tia_phi}
T_{IA} &=& 2iu_\alpha \int d^4x \;\left(
 \phi_\nu^A\partial_\nu\phi^I_\alpha -\phi_\nu^A\partial_\alpha\phi_\nu^I
 +\phi_\alpha^A\partial_\nu\phi_\nu^I 
 -\mu\left(\phi_4^A\phi^I_\alpha -\delta_{\alpha 4}\phi_\nu^A\phi_\nu^I
 +\phi_\alpha^A\phi_4^I\right)\right) ,
\ee
where $\phi^A_\nu=\phi_\nu(x-z)$ and $\phi^I_\nu=\phi_\nu(x)$. From 
(\ref{tia_phi}) we can read off the invariant functions $f_{1,2}$ 
\be
\label{f1_int}
 f_1 &=& 2\int d^4x \,\left\{ \phi_4^A \left(\partial_4\phi_4^I
 + 2\phi_r^I/r + \partial_r\phi_r^I \right) 
 + (\widehat{r{\ts-} z})\cdot\hat r\, \phi_r^A \left( \partial_r\phi_4^I
 - \partial_4\phi_r^I \right) \right. \\
 & & \hspace{2cm}\left.\mbox{} - \mu\left( \phi_4^A \phi_4^I 
  - (\widehat{r{\ts-}z})\cdot\hat r \,\phi_r^A \phi_r^I \right) \right\} , 
  \nonumber \\
\label{f2_int}
 f_2 &=& 2\int d^4x \,\left\{ \hat z\cdot (\widehat{r{\ts-}z})\,\phi_r^A 
  \left(\partial_4\phi_4^I + 2\phi_r^I/r + \partial_r\phi_r^I \right) 
 + \hat r\cdot\hat z\, \phi_4^A \left( \partial_4\phi_r^I
 - \partial_r\phi_4^I \right) \right. \\
 & & \hspace{2cm}\left.\mbox{} -\mu\left(\hat z\cdot\hat r\,\phi_4^A \phi_r^I 
 + \hat z\cdot(\widehat{r{\ts-}z})\, \phi_r^A \phi_4^I \right)  \right\} .
 \nonumber
\ee
The functions $\phi_{r,4}$ are easily determined from (\ref{zm_sol}) 
but the resulting integrals are fairly complicated and have to be 
performed numerically. Results are shown in Fig. \ref{fig_tij} 
which also displays a simple parametrization of the matrix 
elements given in App. \ref{app_tia}. Similar results have been
obtained by Rapp \cite{Rap_97}. The most important point is that 
the asymptotics of $T_{IA}$ are the same as those of the free 
quark propagator. In particular, $T_{IA}\sim\mu^2/x_4$ in the 
time direction and $T_{IA}\sim\mu\sin(\mu r)/r^2$ in the spatial 
direction. 

  In Sect. \ref{sec_zm} we saw that at finite baryon density, quark 
paths tend to wind around the time direction. The probability of an
instanton configuration is proportional to the fermion determinant,
which contains the matrix elements $T_{IA}$. Since these matrix 
elements have the same functional form as the quark propagator, we
also expect the instanton ensemble to be dominated by 
instanton-anti-instanton chains that wind around the time direction.
We will study this phenomenon in Sec. \ref{sec_sim}.

\section{The instanton ensemble at $\mu\neq 0$}
\label{sec_ens}

 The instanton ensemble at finite baryon chemical potential is 
determined by the partition function 
\be
\label{part_fct}
Z &=&  \sum_{N_+,\, N_-} {1 \over N_+ ! N_- !}\int
    \prod_i^{N_+ + N_-} [d\Omega_i\; n(\rho_i) ]
    \exp(-S_{int})\prod_f^{N_f} 
    \det\left(\Dslash+m_f-\mu\gamma_4\right) ,
\ee
Here, $N_+$ and $N_-$ are the numbers of instantons and anti-instantons,
$\Omega_i$ the corresponding collective coordinates (position, color
orientation and size), and $n(\rho)$ the semi-classical size distribution.
The bosonic interaction between instantons is denoted by $S_{int}$
while the fermionic interaction is contained in the determinant
$\det(\Dslash+m-\mu\gamma_4)$. 
  
   Following the usual strategy in instanton model we will split the 
fermion determinant into a high and a low momentum part. The low
momentum part is related to fermion zero modes and will be treated
exactly. For the high momentum part we will assume that it can be
factorized into the contributions of individual instantons. For a
single instanton, the non-zero mode determinant in Gaussian 
approximation is included in the size distribution $n(\rho)$.
The dynamics of the system is then governed by the determinant in
the zero mode basis. At finite chemical potential, the Dirac 
operator has the structure
\be
\label{D_zmz}
\left( i\Dslash+i\mu\gamma_4\right)_{IJ} &=& \left( 
\begin{array}{cc}
 0 & T(\mu) \\ T^\dagger(-\mu) & 0 
\end{array} \right),
\ee
where $T$ is the matrix of overlap matrix elements $T_{IA}$. The 
block off-diagonal form of the Dirac operator is due to the chiral 
structure of the zero modes. Clearly, for $\mu\neq 0$ the Dirac
operator is not hermitean and its eigenvalues are complex.

  For comparison, the form of the Dirac operator in a basis of
$\mu=0$ zero modes is
\be
\label{D_zmz_0}
\left( i\Dslash+i\mu\gamma_4\right)_{IJ} &=& \left( 
\begin{array}{cc}
 0 & T(\mu\!=\!0)+\mu M(\mu\!=\!0) \\ 
 T^\dagger(\mu\!=\!0)-\mu M^\dagger(\mu\!=\!0) & 0 
\end{array} \right) ,
\ee
where $M_{IA}$ are matrix elements of $\gamma_4$. This is the structure
of the Dirac operator that is assumed in random matrix models of QCD
at finite chemical potential (with the additional simplification $M\to 1$)
\cite{Ste_96}. 

   We will study the difference between the exact matrix elements 
(\ref{D_zmz}) and the simplified version (\ref{D_zmz_0}) in Sec. 
\ref{sec_sim}. But even without a calculation it is clear that the 
full matrix elements can carry significant dynamics whereas in the 
simplified form (in particular if the substitution $M\to 1$ is made) 
the chemical potential only acts like an external field. The situation 
is similar to what happens at finite temperature. While random matrix 
models provide a good description of the dynamics of zero modes at $T=0$ 
(and $N_f$ not too large), the dynamics of the phase transition is very 
different (see \cite{WSW_96} for a random matrix model that mimics the
phase transition in the instanton model). There is one more important
difference between the Dirac operators (\ref{D_zmz}) and (\ref{D_zmz_0}):
For $\mu\to\infty$ the schematic Dirac operator (\ref{D_zmz_0}) becomes 
anti-hermitean, so all eigenvalues are imaginary and the determinant
is real. For the exact zero modes, on the other hand, the Dirac operator
does not simplify in that limit.
 
   The statistical system described by the partition function 
(\ref{part_fct}) is quite complicated so that in general one has to 
rely on numerical simulations to determine the phase structure. At 
finite temperature and zero chemical these simulations are straightforward
\cite{SS_96}, but at finite chemical potential one has to deal with 
the fact that the determinant is complex.

   At this point, the only practical solution to this problem is to 
generate an ensemble with the absolute magnitude of the determinant
\be
\label{part_fct_m}
Z_{||} &=&   \sum_{N_+,\, N_-} {1 \over N_+ ! N_- !}
    \int \prod_i^{N_+ + N_-} [d\Omega_i\; n(\rho_i) ]
    \exp(-S_{int})\Big|\prod_f^{N_f} 
    \det\left(\Dslash+m_f-\mu\gamma_4\right)\Big| ,
\ee
and incorporate the phase information into the observables. This means
that expectation values are determined from
\be
\label{vev_ph}
\langle {\cal O}\rangle &=& 
   \frac{\langle e^{i\phi}{\cal O}\rangle_{||}}
	{\langle e^{i\phi}\rangle_{||}},
\ee
where $\langle.\rangle_{||}$ is an expectation value determined with 
the modulus of the determinant and $\phi$ is the phase of the 
determinant. As mentioned in the introduction, taking into account 
the phase information is absolutely essential. The theory described 
by the partition function with the modulus of the determinant is
very different from QCD. In particular, we expect chiral symmetry
restoration to take place at $\mu\simeq m_\pi/2$.

   The problem with this method is that if fluctuations in the 
phase are large then $\langle e^{i\phi}\rangle$ will typically be close 
to zero and the expectation value (\ref{vev_ph}) is essentially 
undetermined. In other words, if fluctuations are big then the 
ensemble generated by (\ref{part_fct_m}) has a very small overlap 
with the correct ground state. 

   An important simplification occurs in the special case of two
colors, $N_c=2$. In this case, the matrix elements $T_{IA}$ are real
(because the color orientation vector $u_\mu$ is real). Eigenvalues 
not only occur in pairs $\pm\lambda$, but also in complex conjugate 
pairs $\lambda,\lambda^*$. Not all eigenvalues are four-fold degenerate
because some of them are purely real or purely imaginary. This means 
that the determinant is real, but not necessarily real and positive. 
We can construct a theory with a real and positive determinant by 
considering the case $N_c=2$ and $N_f$ even.

  Since the determinant is positive, QCD with $N_c=2$ and $N_f=2$ is 
easy to simulate even if the chemical potential is non-zero. But as
in the case of quenched QCD, the physics is very different from real
QCD. If $N_c=2$, baryons are diquarks. Furthermore, the $N_c=2$ theory
has an additional particle-anti-particle (Pauli-G\"ursey) symmetry 
\cite{Pau_57,Gue_58} symmetry which implies that the scalar diquark is
degenerate with the pion. This means that the critical chemical 
potential is again expected to be $\mu\simeq m_\pi/2$.

\section{Spectrum of the Dirac operator at finite chemical potential}
\label{sec_spec}

  The low virtuality part of the Dirac spectrum is intimately 
connected with the physics of chiral symmetry breaking. For $\mu=0$,
this is expressed by the Banks-Casher relation
\be 
\label{bc}
\langle \bar qq\rangle &=& -\pi\rho(\lambda\!=\!0),
\ee
which relates the quark condensate in the chiral limit to the 
spectral density of the Dirac operator, $\rho(\lambda)=\langle
\sum_i \delta(\lambda-\lambda_i)\rangle$. For non-zero chemical
potential, the eigenvalues are complex and $\rho(\lambda)$ is
a density in the complex plane. How to interpret the Banks-Casher 
relation in that case was explained by Stephanov \cite{Ste_96}:
The quark condensate is related to the resolvent of the Dirac
operator, $\langle \bar qq\rangle =G(i0)$ where $G(z)=\langle
{\rm tr}(i\Dslash+z)^{-1}\rangle$ defines a function in the 
complex plane. The vector field $\vec G=({\rm Re}G,-{\rm Im}G)$, 
can be interpreted as the ``electric field" generated by the 
``charge density" $\rho(z)$. If quarks are condensed this 
function has a discontinuity on the real line. This means 
that there has to be a finite density of eigenvalues on that
line. 

  If the chemical potential is smaller than the critical value 
where the baryon density becomes non-zero the quark condensate
has to be independent of $\mu$. This means that while in general
eigenvalues will spread in the complex plane, eigenvalues near
zero virtuality have to stay on the real line. As soon as $\mu>
\mu_c$, some of the small eigenvalues can move into the complex
plane, but the density of eigenvalues on the real line has to
remain finite. Only above the critical chemical potential for
chiral symmetry restoration can all eigenvalues spread out in
the complex plane. 

  In the remainder of this section we will discuss eigenvalue 
distributions in the instanton liquid model. These distributions
were obtained from numerical simulations of the partition function
(\ref{part_fct_m}). For simplicity, we have kept the instanton 
density $(N/V)=1\,{\rm fm}^{-4}$ fixed. In principle, the correct 
equilibrium density is determined by minimizing the free energy.
This can be done using the methods discussed in \cite{SS_96}, but 
is beyond the scope of our exploratory study. 

  Fig. \ref{fig_mu_spec} shows scatterplots of eigenvalues
of the Dirac operator obtained in quenched simulations for 
$N_c=2,3$ and $\mu=1\,{\rm fm}^{-1}$. For $N_c=3$, the eigenvalues
spread almost uniformly over a band in the complex plane. In the
case of two colors, there is a finite fraction of eigenvalues 
that stays on the real line. This phenomenon was also observed
in random matrix models \cite{HOV_97}. From our discussion above 
that would seem to indicate that for $N_c=2$ chiral symmetry remains
broken, while for $N_c=3$ chiral symmetry is restored. This is
not the case. Chiral symmetry restoration in the case of three
colors is an artefact of the quenched approximation. For $N_c=2$ 
the density of eigenvalues on the real line vanishes in the 
infinite volume limit. We will study this case in more detail 
in the next section.

  For $N_c=2$ the unquenched eigenvalue distribution is 
qualitatively similar to the quenched distribution. This is
not true for $N_c=3$. In order to calculate the unquenched 
eigenvalue distribution we have binned the eigenvalues and 
determined the average number of eigenvalues per bin from 
numerical simulations of the partition function (\ref{part_fct_m}). 
The unquenched distribution is then obtained by including
the phase of the determinant according to (\ref{vev_ph}).
In general, this procedure will give a complex eigenvalue 
density but in our simulations the imaginary part was 
always consistent with zero.

  In Fig. \ref{fig_mu_cont} we show the density of eigenvalues
for $\mu=1.0\,{\rm fm}^{-1}$ without the phase (upper panel) and 
with the phase included (lower panel). We observe that if the phase 
is included, the number of eigenvalues away from the real line
near $\lambda=0$ is reduced. This is clearly the correct tendency
although with our statistics, we cannot really determine whether 
there is a discontinuity on the real line. We will study the
quark condensate more quantitatively in the next section. 

\section{Chiral symmetry restoration in the instanton liquid}
\label{sec_sim}

  In this section we wish to study the quark condensate in 
unquenched simulations of the instanton liquid at finite 
chemical potential. Let us start with the simpler case 
$N_c=2,\, N_f=2$. In Fig. \ref{fig_qq_nc2} we show the 
quark condensate and the fraction of real eigenvalues as 
a function of the chemical potential for different volumes
$V=16,32,64\,{\rm fm}^4$. The instanton density was fixed
at $(N/V)=1\,{\rm fm}^{-4}$ and the quark mass is $m=20$ MeV.
This mass is somewhat larger than the physical quark mass 
and was chosen to minimize finite volume effects. We observe 
that the quark condensate decreases as $\mu$ increases and that 
chiral symmetry is restored for large chemical potentials. 
While the number of real eigenvalues decreases as the volume is
increased, the behavior of the quark condensate does not change 
appreciably. This would seem to indicate that for finite quark
masses, there is no sharp transition in the infinite volume limit.

  The behavior of the quark condensate and the chiral susceptibility
$\chi_\sigma = V (\langle(\bar qq)^2\rangle -\langle\bar qq\rangle^2)$
for different quark masses $m=20,30,40$ MeV is shown in Fig. 
\ref{fig_sus_nc2}. We have also determined the $\mu=0$ pion mass
for these quark masses. We find $m_\pi\simeq 240,280,360$ MeV,
with statistical errors $\Delta m_\pi\simeq 20$ MeV. The onset 
$\mu_0$ of chiral symmetry restoration is hard to determine from 
the data, but the results appear to be consistent with $\mu_0
\simeq m_\pi/2$. We also observe that the pseudo-critical chemical 
potential $\mu_c$ (determined from the peak of the chiral susceptibility) 
drops with the quark mass. Again the errors are large, but the 
behavior is compatible with $\mu_c\sim \sqrt{m}$.

  For large chemical potential the quark condensate goes to zero
but we expect that diquarks are condensed. This idea is based on 
the fact that $SU(2)$ gauge theory with two massless flavors has 
an extra symmetry that relates the chiral condensate $\langle \bar 
q^aq^a\rangle$ to the scalar diquark condensate $\langle\epsilon^{ab}
q^aC\gamma_5\tau_2 q^b\rangle$. Here, $C$ is the charge conjugation 
matrix, $a,b$ are color labels and $\tau_2$ is the anti-symmetric 
Pauli matrix (the diquark is an iso-singlet). A finite quark mass
breaks the symmetry and favors quark-anti-quark over diquark 
condensation. For large chemical potential, on the other hand, 
we expect that diquark condensation is favored. This scenario was
first discussed in the context of the strong coupling limit by 
Dagotto et al. \cite{DKM_86}. However, these authors concluded 
that chiral symmetry remains broken for large $\mu$. In general, 
this is not correct. For $N_f=2$ the diquark condensate is a chiral 
singlet. For more than two flavors, on the other hand, part of the 
chiral symmetry is broken\footnote{Dagotto et al.~use staggered 
fermions, so effectively the number of flavors is a multiple of 
4}. 

  Let us now study the more interesting case $N_c=3$. We have performed 
simulations with the Dirac operator evaluated in a basis of exact $\mu
\neq 0$ zero modes (\ref{D_zmz}) and with the schematic Dirac operator 
(\ref{D_zmz_0}). The results presented here were obtained by generating 
5000 configurations in a volume $V=16\,{\rm fm}^4$. In addition to that,
some runs were performed with $V=32\,{\rm fm}^4$. To illustrate the 
procedure, we show trajectories of the quark condensate and the phase 
of the determinant for 400 Monte Carlo sweeps at $\mu=0.5\,{\rm fm}^{-1}$ 
in Fig. \ref{fig_traj}. The fluctuations are larger as compared to the 
$\mu=0$ case, but $\langle e^{i\phi}\rangle$ is still differs significantly 
from zero. If $\mu>1\,{\rm fm}^{-1}$ the fluctuations are so large that
both the real and imaginary parts of $\langle e^{i\phi}\rangle$ are
close to zero.

  Fig. \ref{fig_qq_nc3_0} shows the real parts of the quark condensate 
and the average phase $\langle e^{i\phi}\rangle$ calculated with the 
$\mu=0$ zero modes. The imaginary part of both quantities is always 
consistent with zero. We display both $\langle\bar qq\rangle$ and 
$\langle\bar qq\rangle_{||}$, the quark condensate with and without
fluctuations of the phase of the determinant taken into account. 
    
  We observe that $\langle\bar qq\rangle_{||}$ drops rapidly with $\mu$. 
This is the expected artefact that comes from neglecting the phase of 
the determinant. With the phase taken into account, $\langle\bar qq
\rangle$ becomes almost independent of the chemical potential for
$\mu<1\,{\rm fm}^{-1}$. In general we expect $\langle \bar qq\rangle$ 
to be $\mu$-independent for $\mu<\mu_c$ with $\mu_c\simeq m_B/3
\simeq 1.5\,{\rm fm}^{-1}$. In the instanton model this is not quite
true, because the most tightly bound state that carries baryon 
number is a scalar diquark, so $\mu_c\simeq m_D/2$.

  The data show a small but statistically significant rise in 
$\langle\bar qq\rangle$ for $\mu<1\,{\rm fm}^{-1}$. The result is
not a finite volume effect because it is also present in the $V=32
\,{\rm fm}^4$ data. A similar phenomenon was also observed in 
random matrix models \cite{HJV_97}, where it was argued that the 
problem is related to the restriction to a finite number of 
static modes. 

    For chemical potentials
larger than $\mu\simeq 1\,{\rm fm}^{-1}$ the complex phase becomes
very small and $\langle\bar qq\rangle$ has very large error bars. 
Without the phase, the quark condensate is small and fluctuates
little. This is due to the fact that the eigenvalues of the Dirac
operator are pushed farther and farther away from the real line,
see Fig. \ref{fig_mu_spec}b. If the chemical potential is very 
large, $\mu> 5\,{\rm fm}^{-1}$, the complex phase factor 
starts to grow and $\langle \bar qq\rangle$ can again be determined. 
The reason for this behavior is trivial and was already discussed
in Sect. \ref{sec_ens}: If $\mu M_{IA}$ is much bigger than the
typical overlap matrix element $T_{IA}$, the Dirac operator 
becomes anti-hermitean and all eigenvalues are close to the 
imaginary line\footnote{Note that we have performed these 
simulations without the exponential suppression factor 
(\ref{dens_mu}). This was done for consistency, because
(\ref{D_zmz_0}) is really a small $\mu$ approximation. 
Including the suppression factor also has the unusual effect
that $\mu M_{IA}$ never dominates over $T_{IA}$, because
$T_{IA}\sim\rho^{-1}$ and the factor $\exp(-N_f(\mu\rho)^2)$
in the instanton rate forces $\rho^{-1}>\mu$.}.

 In Fig. \ref{fig_qq_nc3_1} we show the results for the Dirac
operator evaluated in the basis of exact $\mu\neq 0$ zero modes.
Again, $\langle\bar qq\rangle_{||}$ drops rapidly with $\mu$. 
Including the phase helps to stabilize the quark condensate, 
but there is a definite drop in $\langle\bar qq\rangle$ even
for $\mu<1\,{\rm fm}^{-1}$. In this case, the effect becomes
smaller as we go to larger volumes. We therefore believe that
the effect is mostly due to the onset of chiral symmetry 
restoration near $\mu_c\simeq 1\,{\rm fm}^{-1}$, smeared out
by finite volume effects. 

  For $\mu> 1\,{\rm fm}^{-1}$ the complex phase factor
again becomes very small and we cannot perform reliable 
simulations. However, for $\mu>3\,{\rm fm}^{-1}$ the phase
factor $\langle e^{i\phi}\rangle$ starts to grow and simulations 
are feasible. In the plasma phase, chemical potentials between
$(2-3)\,{\rm fm}^{-1}$ correspond to a baryon density $n_B=
2\mu^3/(3\pi^2)=(0.5-1.8)\,{\rm fm}^{-3}$. In this regime the 
quark condensate is very small and the instanton ensemble is 
in the chirally restored phase. It is interesting to analyze 
the mechanism that causes the symmetry to be restored. 

  For this purpose we show two typical instanton configurations
for $\mu=0$ and $\mu=5\,{\rm fm}^{-1}$ in Fig. \ref{fig_muconf}. 
The figures show projections of a four dimensional box into the 
$x_3-x_4$ plane. The locations of instantons and anti-instantons 
are denoted by $\pm$ signs, and the strength of the fermionic 
overlap matrix elements $T_{IA}$ are indicated by the thickness 
of the lines connecting the instantons. We observe that for $\mu=0$ 
instantons are distributed randomly and there is no particular pattern 
to the lines connecting them. This is a sign for chiral symmetry 
breaking. If instantons tend to form large random clusters, then 
the corresponding eigenfunctions will be low-lying delocalized modes. 
These are the modes that form the quark condensate. 

   At large chemical potential, instantons tend to line up in the 
$x_4$-direction. These chains are weakly linked in the spatial
direction. The reason why these configurations are favored should
be clear from our discussion of the overlap matrix elements in 
Sect. \ref{sec_tia}: For large $\mu$, $T_{IA}$ is big in the 
temporal direction and oscillates for spacelike separations. 
There are two reasons why these configurations tend to have 
an almost real determinant. One is the fact that for the 
most attractive orientation, we have $u_4=1$ and the overlap
matrix element is real. The other is that the dominant terms
in $T_{IA}$ at large chemical potential are even in $\mu$,
corresponding to an almost hermitean Dirac operator. 

  For the chain-like configurations observed at large chemical
potential, chiral symmetry is restored because the eigenfunctions
tend to be spatially localized. This can be seen from Fig. 
\ref{fig_part} which shows the distribution of participation 
numbers of the low-lying eigenstates for three different 
simulations at $T=\mu=0$, $T\simeq T_c$ and $\mu=0$ \cite{SS_96},
and $T=0$, $\mu>\mu_c$. The participation number is a concept
borrowed from the theory of localization (Mott-Anderson) phase 
transitions \cite{KM_93}. For a normalized eigenvector $c_i$, we 
define the participation number
\be
\label{part_num}
P = 1/\sum_{i} |c_i|^4 \, \hspace{1cm}
\left( \sum_i |c_i|^2=1 \right).
\ee
The participation number is a measure of the number of (localized)
basis states that contribute to an eigenstate of the system. We 
observe that at $\mu=T=0$ many basis states participate in the 
low-lying eigenstates. In fact, the participation number of these
states grows with the volume of the system. The finite $T$ transition
is associated with the formation of instanton-anti-instanton 
molecules and the dominant participation number near $T_c$ is 
two. In the finite $\mu$ transition we also observe that the 
average participation number becomes smaller, but the average 
cluster size is bigger than two. One can check that the size
of these clusters is dominantly controlled by the inverse 
temperature (the length of the imaginary time direction). We
conclude that the system is dominated by oriented instanton
``polymers". 

  In more physical terms, chiral symmetry restoration is due 
the presence of valence quarks in the system. Valence quarks can 
saturate the fermionic bonds between instantons. If the density of 
valence quarks is sufficiently large all bonds are saturated, and 
additional quarks can no longer propagate from one instanton to 
another. As a result, quarks do not acquire a constituent mass 
and chiral symmetry is restored. This interpretation becomes 
more obvious if one reduces the instanton-induced interaction 
between quarks to an effective four-fermion interaction, and 
treats this interaction in the mean-field approximation (see,
e.g. \cite{KLW_90}). Chiral symmetry breaking is then described
in terms of a gap equation for the constituent quark mass. At
large density chiral symmetry is restored because valence quarks
Pauli-block the quark loop that drives the formation of a 
constituent mass.

\section{Summary}
\label{sec_sum}

 In summary we have studied instantons and their interactions at 
finite baryon chemical potential. We have emphasized the importance 
of exact zero modes in the spectrum of the Dirac operator even 
at $\mu\neq 0$. We have determined matrix elements of the Dirac
operator between zero modes associated with different instantons.
These matrix elements favor the propagation of quarks along the 
positive $x_4$ direction. 

 Using these ingredients we have performed a number of exploratory
simulations of the instanton liquid at finite chemical potential.
In the case $N_c=2$ and $N_f=2$ these simulations are straightforward
because the theory does not have sign problem. We observe that chiral
symmetry is restored at $\mu_c\sim \sqrt{m}$. The high density state
is likely to support a diquark condensate. 

 For $N_c=3$ the simulations are complicated by the sign problem. In
our simulations we focus on the dynamics of (quasi) zero modes. This
means that number of degrees of freedom in a given volume is much 
smaller than in a typical lattice simulation and the sign problem 
is not as severe. In practice we can perform simulations in 
sufficiently large volumes for $\mu< 1\,{\rm fm}^{-1}$ and 
$\mu> 3\,{\rm fm}^{-1}$. For small chemical potential we 
observe that including the phase of the determinant has the 
correct effect of stabilizing the quark condensate. Qualitatively,
this is true for both the ensemble generated with the $\mu=0$
and the correct $\mu\neq 0$ zero modes. Quantitatively, the 
quark condensate rises for small $\mu$ in the first case, while
there is a tendency towards chiral restoration in the other case. 

   The structure of the chirally restored phase at large $\mu$ is 
completely determined by the functional form of the zero
mode wave functions. In schematic models of the structure of the
Dirac operator, the fluctuations in the phase become small because
for large $\mu$ the determinant is dominated by the chemical potential
term. Chiral symmetry is restored because the large external field 
pushes all eigenvalues away from zero. In the instanton liquid, 
fluctuations become small because the ensemble is dominated by 
chain-like configurations for which the determinant is almost 
real. Chiral symmetry is restored because the corresponding
eigenfunctions are strongly localized. 

  Much more work remains to be done in order to understand the 
nature of the chirally restored phase at large chemical potential
and the onset of chiral restoration at small $\mu$. In particular,
we have to study the quark propagator, mesonic correlation functions
and the thermodynamics of the system at large $\mu$. This should
clarify whether there is any tendency towards diquark condensation,
strange matter formation, etc.

  We should also point out the shortcomings of our procedure. The
restriction of the determinant to quasi zero modes has enabled us
to perform simulations in reasonably large volumes but it also 
introduces certain artefacts. For example, non-zero modes have to 
contribute to the total baryon number of the system. Also, if 
non-zero modes are neglected the quark condensate is not necessarily
independent of $\mu$ below $\mu_c\simeq m_B/3$. Finally, as long 
as there is no real solution to the sign problem, simulations 
in the interesting regime $\rho_B\simeq \rho_B^0$ will remain
impossible. 

 Acknowledgements: I would like to thank R. Rapp, E. Shuryak, M. 
Velkovsky and J. Verbaarschot for useful discussions. 

\newpage

\appendix
\section{Fermionic overlap matrix elements}
\label{app_tia}

  The two functions $f_{1,2}$ depend on the instanton-anti-instanton
distance $(\vec z,z_4)$, the two radii $\rho_{I,A}$ and the chemical
potential $\mu$. If the two radii are equal, it is clear that $f_i$
only depends on $|\vec z|/\rho$, $z_4/\rho$ and $\mu\rho$. We have 
checked numerically that if $\rho_I\neq\rho_A$ the matrix elements 
depend to a very good accuracy only on the geometric mean $\overline
\rho=\sqrt{\rho_I\rho_A}$. 

  This leaves us with two functions $f_i(\tilde z_r,\tilde z_4,\tilde
\mu)$, where $\tilde z_r=z_r/\overline\rho$, etc. are dimensionless 
variables. We have determined these functions numerically for many 
different values of the variables and fitted the result to the 
following parametrizations
\be
\label{f1_par}
 f_1 &=& \frac{2.0}{(2.2+z^2)^2}  \Big\{
   \left( 2 z_4 + a_1 \mu z_4^2
     + a_2 \mu z_r^2 + a_3 \mu^2 z_4 \right) \cos(\mu z_r) \\
     & &  \hspace{1cm} \mbox{}
     - \left( a_4 \mu^2 + a_5 z_r^2 - a_6 z_4^2
	- a_7 \mu z_4 z2 \right) \frac{\sin(\mu z_r)}{z_r} 
	+ a_8 \mu^2 \sin(\mu z_r) \Big\}, \nonumber \\
\label{f2_par}
 f_2 &=& \frac{2.0}{(2.2+z^2)^2}   \Big\{ 
    \left( 2 z_r - b_1 \mu z_4^3/z_r + b_2 \mu z_4 z_r 
      + b_3 \mu^2 z_r^2 \right) \cos(\mu z_r)   \\
     & &  \hspace{1cm} \mbox{}
      + \left( b_1 z_4^3 + 3 b_4 z_r^2 z_4  
      + b_5 \mu z_r^4 + b_6 \mu z_r^2 z_4^2 \right)
	  \frac{\sin(\mu z_r)}{z_r^2} 
      + \left(b_7 \mu + b_8 \mu^2 z_4\right) \sin(\mu z_r) 
       \Big\}, \nonumber 
\ee
where we have dropped the tilde symbol. The functional form of 
the parametrizations was inspired by the free quark propagator 
(\ref{s_mu}). In addition to that we have to respect the symmetries 
$f_i(z,\mu)=\mp f_i(-z,-\mu)$, ($i$=1,2) of the matrix elements 
and ensure the correct behavior for $\mu\to 0$. 

  Fitting the parametrization (\ref{f1_par},\ref{f2_par}) to the 
data, we get
\be
\begin{array}{lll}
 a_1 =    1.334, \hspace{0.5cm} &
 a_2 =    2.142, \hspace{0.5cm} & 
 a_3 =    2.872, \\    
 a_4 =    0.756, &     
 a_5 =    2.378, &   
 a_6 =    0.893, \\    
 a_7 =    1.060, &    
 a_8 =   -0.845, &
\end{array}\\
\begin{array}{lll}
 b_1 =    2.361, \hspace{0.5cm} &   
 b_2 =    1.561, \hspace{0.5cm} & 
 b_3 =    0.072, \\
 b_4 =    1.027, &
 b_5 =    0.818, &
 b_6 =    0.844, \\
 b_7 =    3.270, &
 b_8 =    1.313. &
\end{array}
\ee
The quality of the fit can be judged from the results shown in 
Fig. \ref{fig_tij} and appears quite satisfactory for our purposes.

\newpage
\bibliography{[schaefer.tex.revrev]rev}


\newpage\noindent

\begin{figure}
\caption{\label{fig_tij}
Overlap matrix elements $f_{1,2}$ as a function of the 
instanton-antiinstanton separation $z$ in units of $\rho$.
Fig. a) shows $f_1$ as a function of $z_4$ for $z_r=0$, 
while Fig. b) shows $f_2$ as a function of $z_r$ for $z_4=0$.
The points are numerical results for the overlap integrals  
while the lines show the parametrization discussed in the 
appendix. The solid, long-dashed, and short-dashed curves 
correspond to $(\mu\rho)=0,1,2$.}
\end{figure} 

\begin{figure}
\caption{\label{fig_mu_spec}
Scatterplots of the complex eigenvalues of the Dirac operator from
quenched simulations at $\mu=1\,{\rm fm}^{-1}$ for $N_c=2$ and $N_c=3$.}
\end{figure} 

\begin{figure}
\caption{\label{fig_mu_cont}
Contourplots of the eigenvalue density in the complex plane from 
unquenched simulations at $\mu=1\,{\rm fm}^{-1}$ without (upper
panel) and with (lower panel) the phase of the determinant included.
The $x$ and $y$ axes show the real and imaginary parts of the 
eigenvalues.}
\end{figure} 

\begin{figure}
\caption{\label{fig_qq_nc2}
Quark condensate (upper panel) and fraction of real eigenvalues (lower
panel) as a function of the chemical potential for $N_c=2,\,N_f=2$ from 
numerical simulation of the unquenched theory. The different curves 
correspond to different volumes $V=16,32,64\,{\rm fm}^4$.}
\end{figure} 

\begin{figure}
\caption{\label{fig_sus_nc2}
Quark condensate (upper panel) and scalar susceptibility (lower panel) 
as a function of the chemical potential in $N_c=2,\,N_f=2$ QCD. The 
different curves show the results for different quark masses $m=10,
20$ MeV with the volume $V=32\,{\rm fm}^4$ kept fixed.}
\end{figure} 

\begin{figure}
\caption{\label{fig_traj}
Time history of the quark condensate (both real and imaginary parts) 
and the cosine of the phase of the fermion determinant for an 
instanton liquid simulation at $\mu=0.5\,{\rm fm}^{-1}$. The 
figure shows 400 configurations generated in a volume $V=16\,
{\rm fm}^4$.}
\end{figure} 

\begin{figure}
\caption{\label{fig_qq_nc3_0}
Quark condensate (upper panels) and average phase of the fermion determinant 
(lower panels) as a function of the chemical potential for $N_c=3,\,N_f=2$ 
from numerical simulation of the unquenched theory. The open and solid
points show the quark condensate without and with the phase 
included. The calculations shown here were performed in the $\mu=0$
basis equ. (\ref{D_zmz_0}). }
\end{figure} 

\begin{figure}
\caption{\label{fig_qq_nc3_1}
Same as Fig. (\ref{fig_qq_nc3_0}) with the determinant evaluated in 
the $\mu\neq 0$ basis (\ref{D_zmz}).}
\end{figure} 

\begin{figure}
\caption{\label{fig_muconf}
Typical instanton configurations for $\mu=0$ and $\mu=5\,{\rm fm}^{-1}$.
The plots show projections of a four dimensional box into the $x_3-x_4$
plane. The positions of instantons and anti-instantons are shown as
$\pm$ symbols. The lines indicate the strength of the fermionic overlap
matrix elements $T_{IA}$.}
\end{figure} 

\begin{figure}
\caption{\label{fig_part}
Distribution of participation numbers for low-lying eigenvectors in the
instanton liquid at finite temperature (upper panel), and at finite 
chemical potential (lower panel).}
\end{figure}

\pagestyle{empty}

\newpage
\setcounter{figure}{0}

\begin{figure}
\begin{center}
\leavevmode
\epsfxsize=14cm
\epsffile{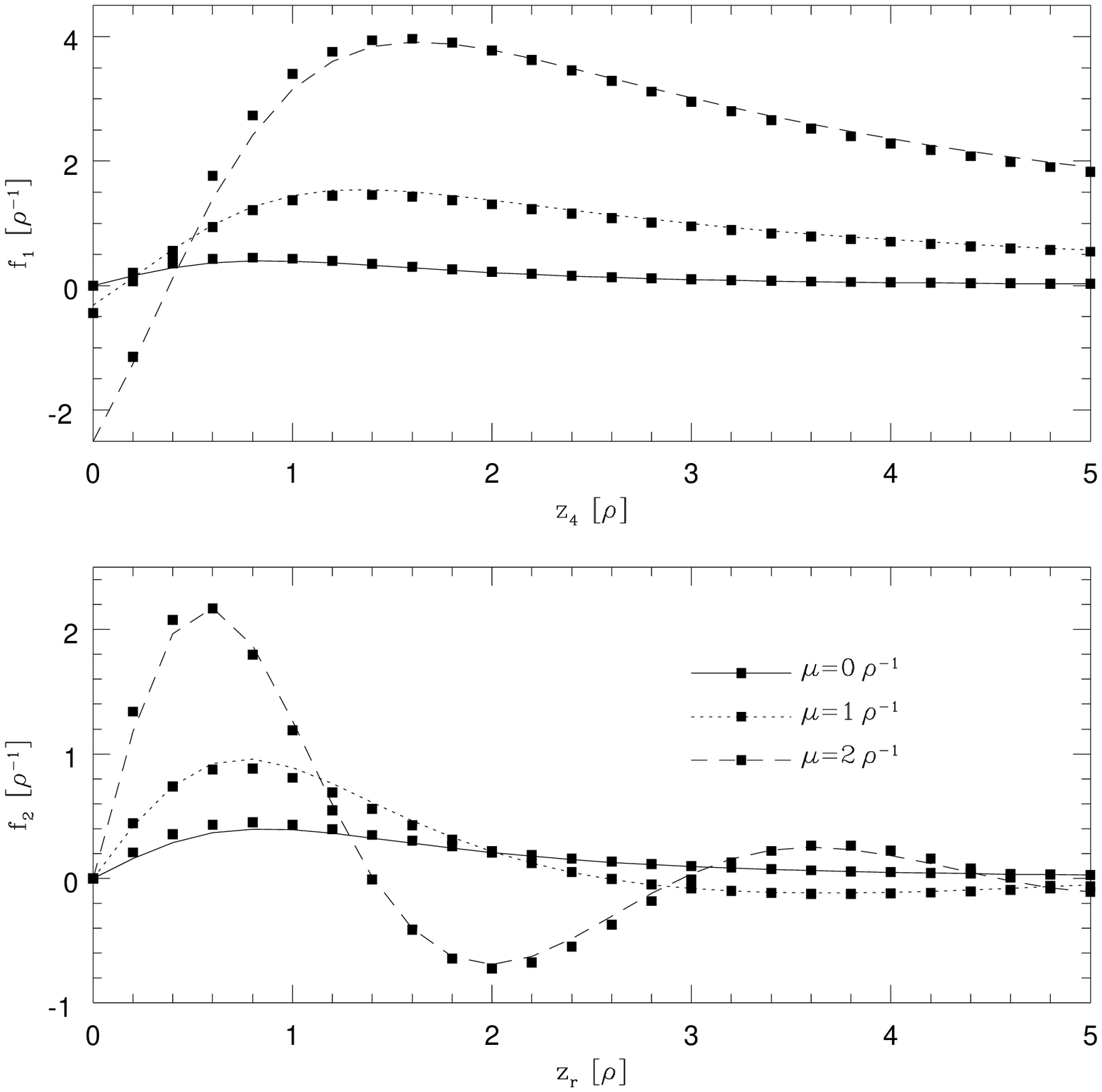}
\end{center}
\caption{}
\end{figure}
\vfill

\begin{figure}
\begin{center}
\leavevmode
\epsfxsize=8cm
\epsffile{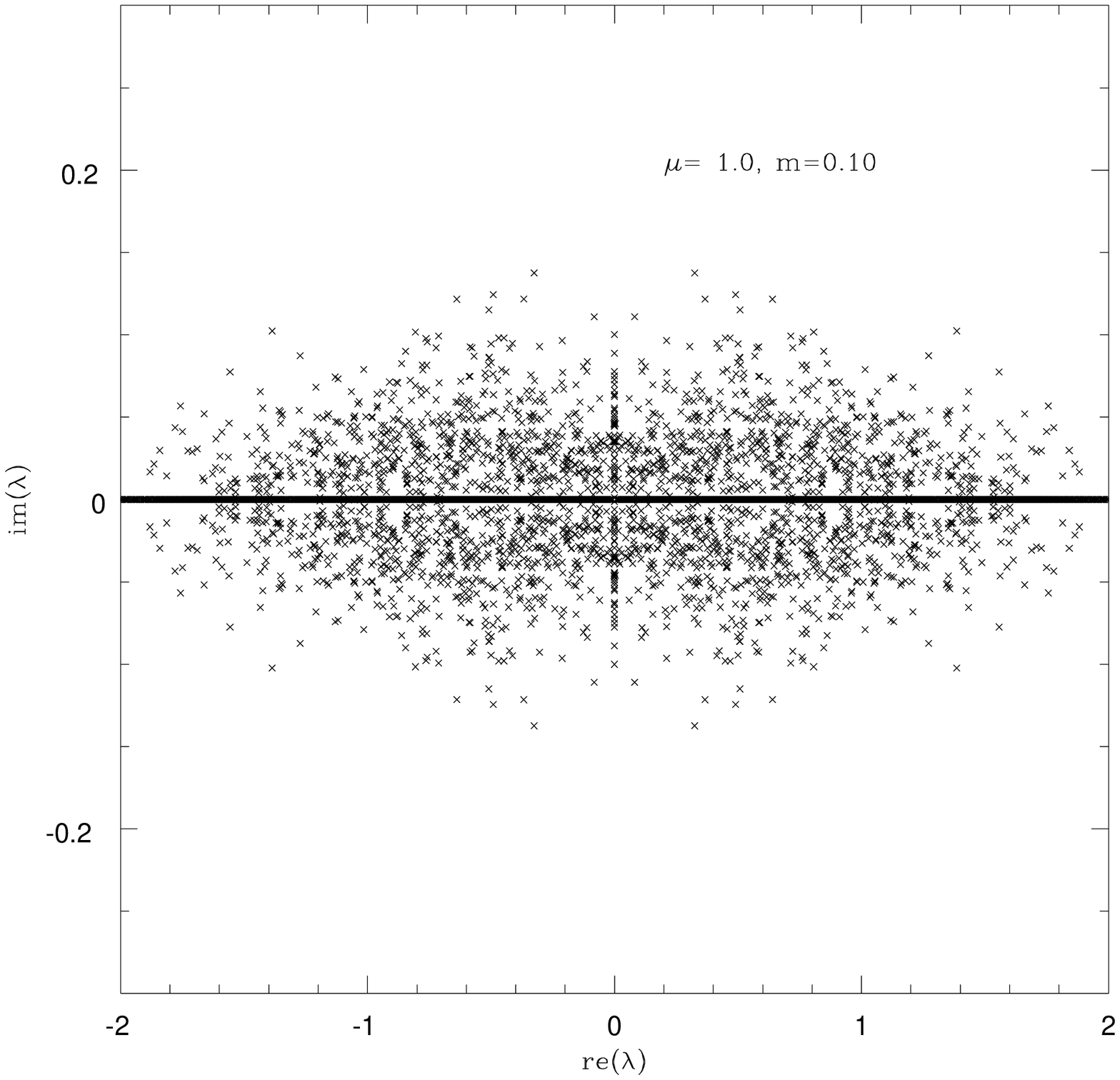}
\end{center}
\begin{center}
\leavevmode
\epsfxsize=8cm
\epsffile{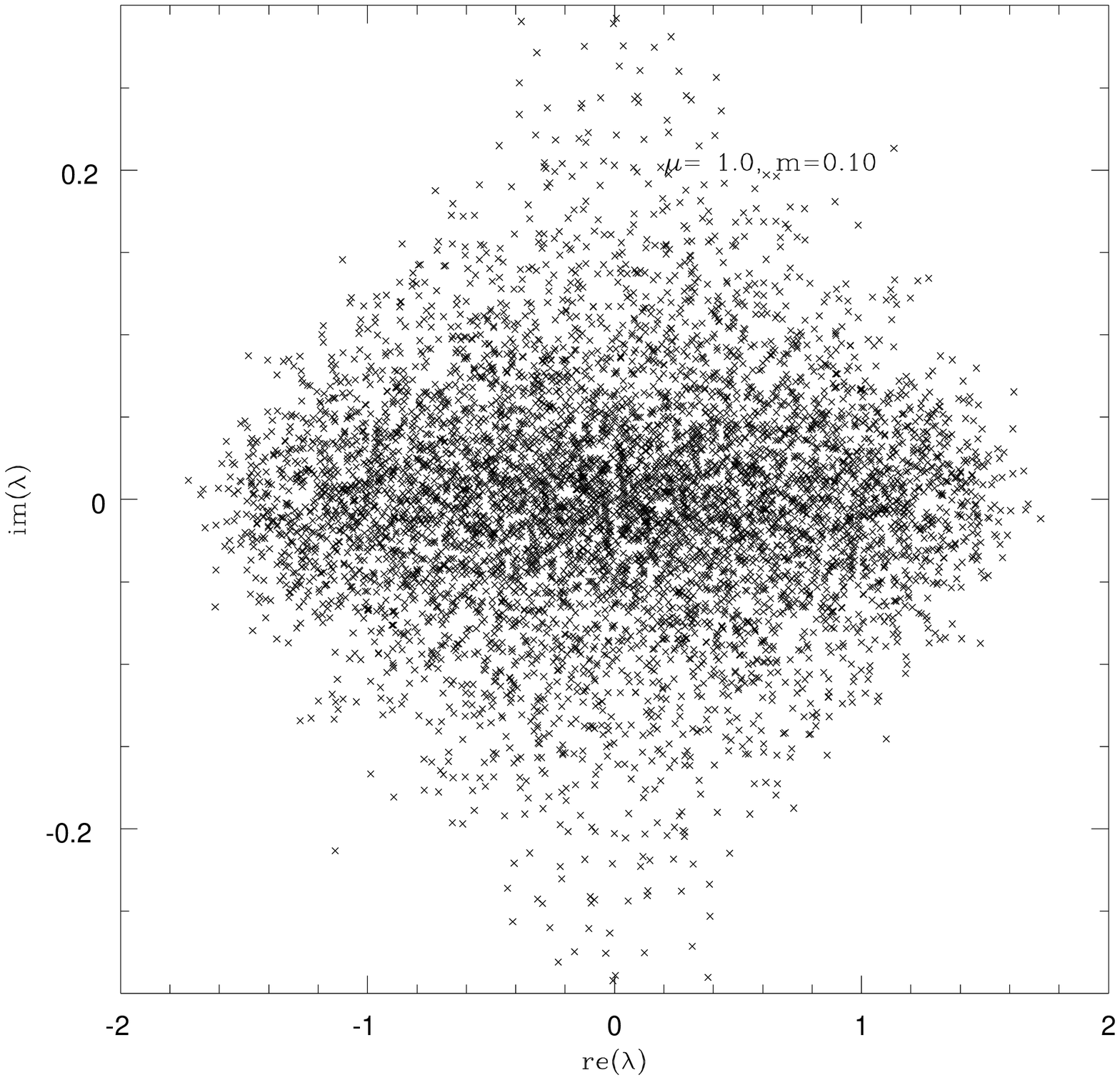}
\end{center}
\caption{}
\end{figure}
\vfill

\begin{figure}
\begin{center}
\leavevmode
\epsfxsize=8cm
\epsffile{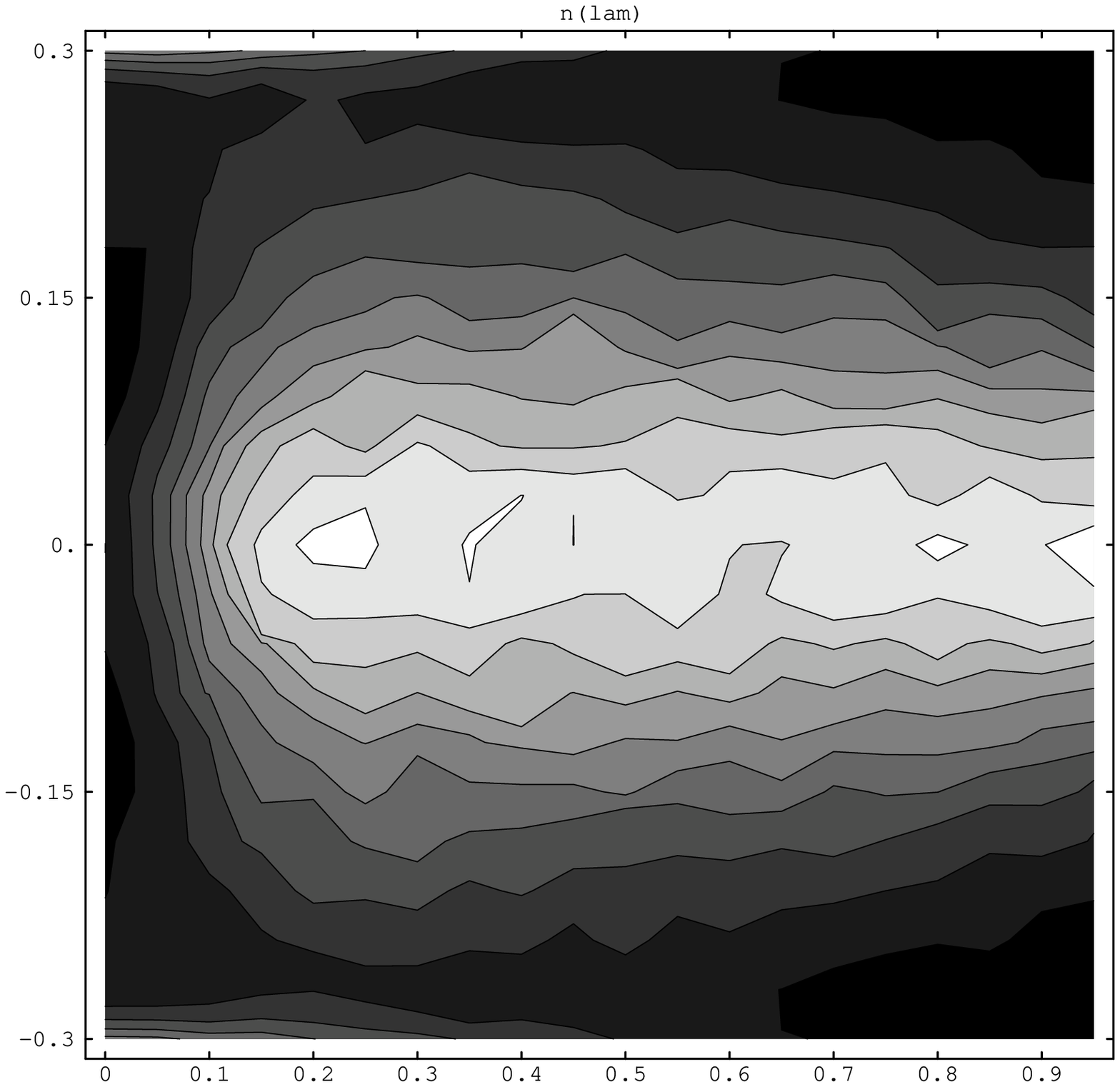}
\end{center}
\vspace*{-1cm}
\begin{center}
\leavevmode
\epsfxsize=8cm
\epsffile{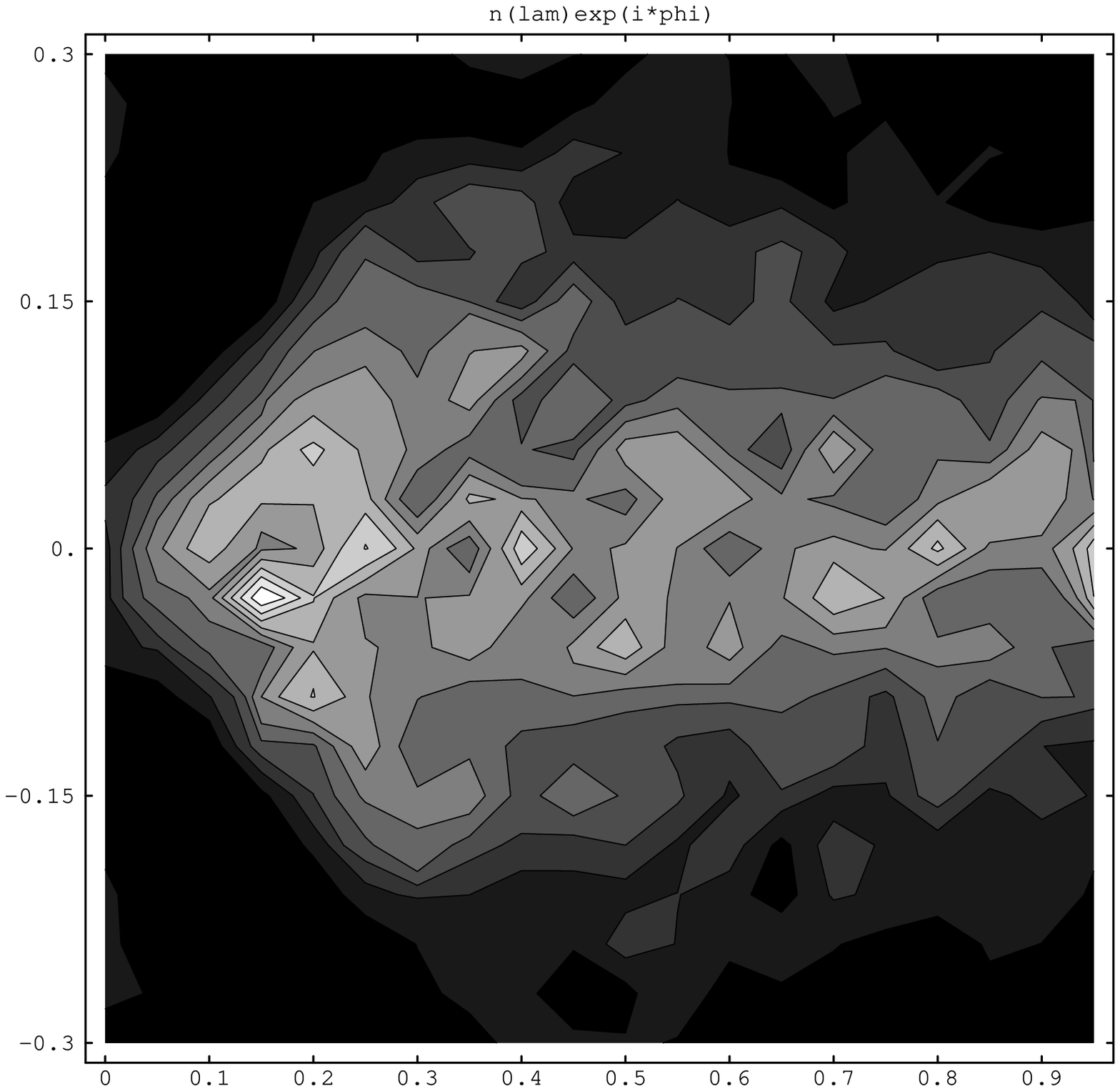}
\end{center}
\caption{}
\end{figure}
\vfill

\begin{figure}
\begin{center}
\leavevmode
\epsfxsize=14cm
\epsffile{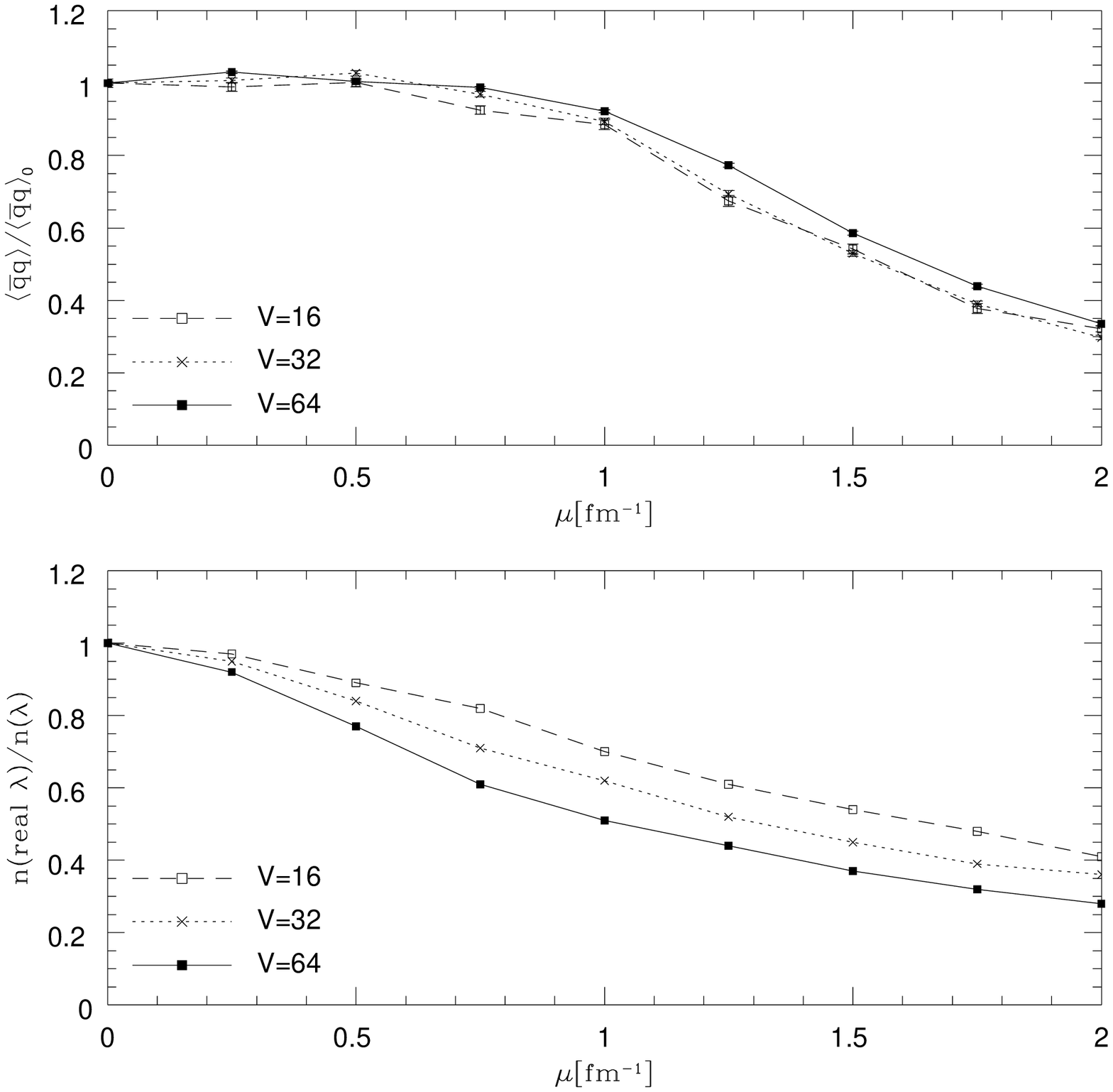}
\end{center}
\caption{}
\end{figure}
\vfill

\begin{figure}
\begin{center}
\leavevmode
\epsfxsize=14cm
\epsffile{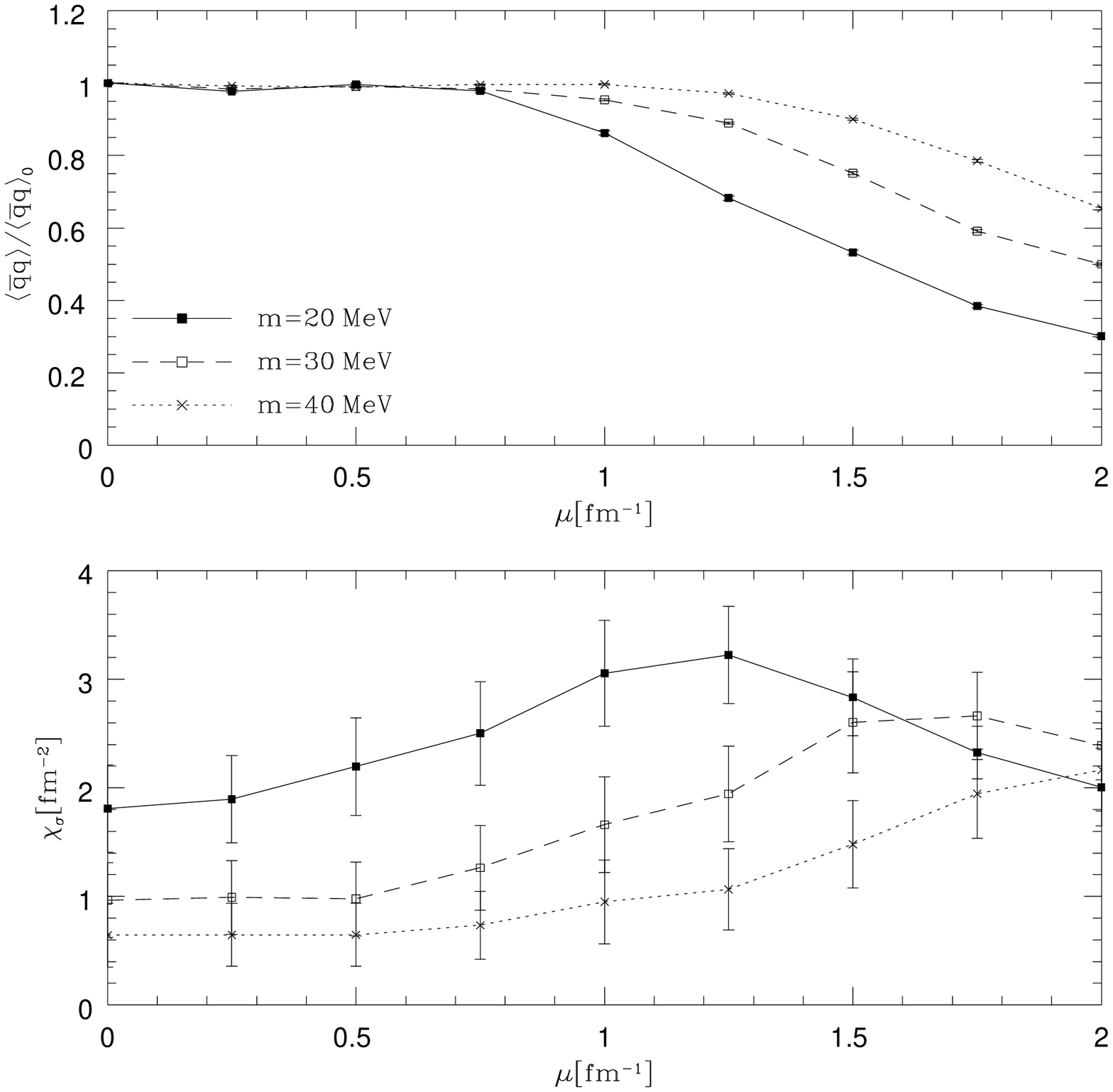}
\end{center}
\caption{}
\end{figure}
\vfill
\newpage

\begin{figure}
\begin{center}
\leavevmode
\epsfxsize=14cm
\epsffile{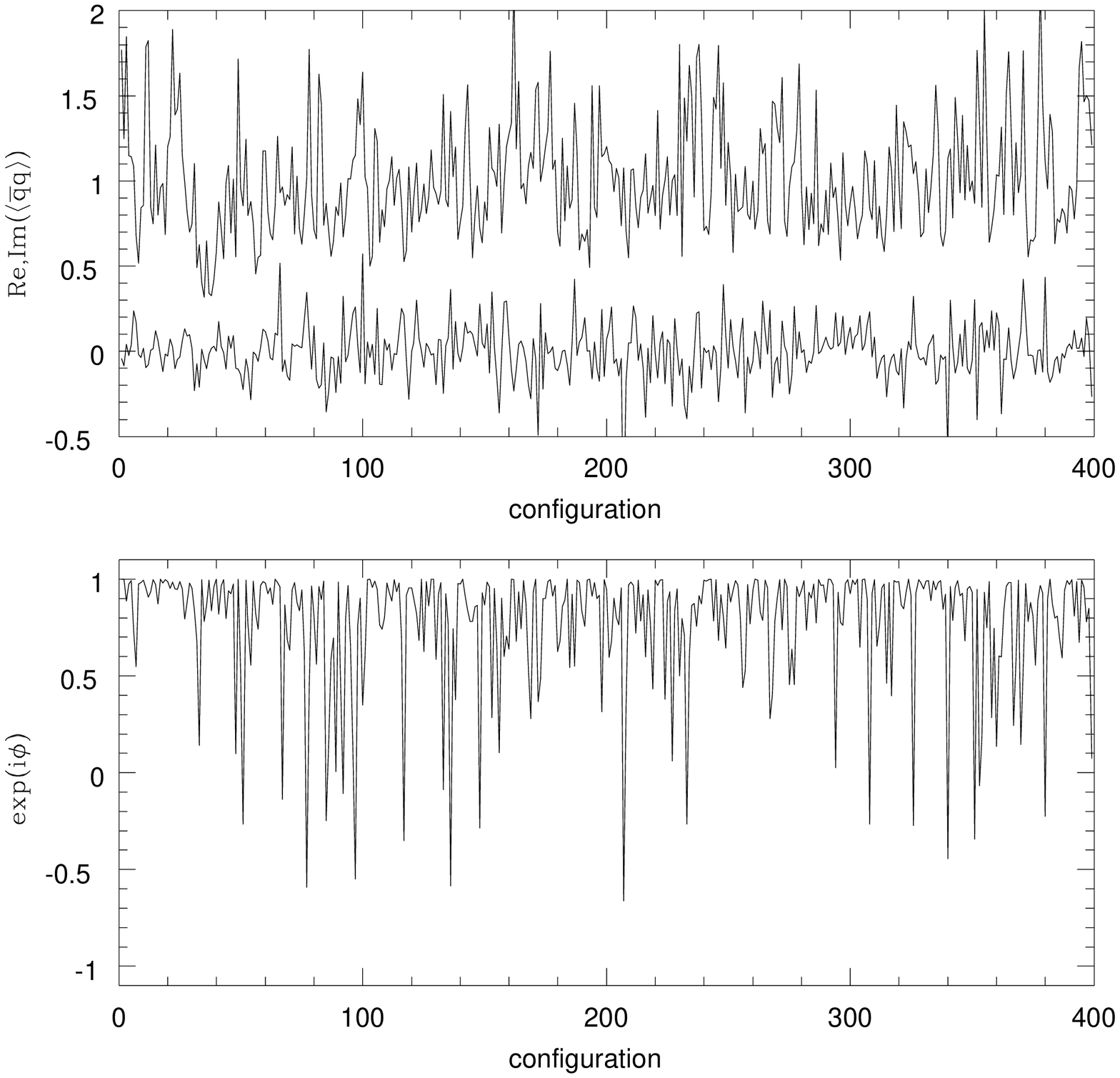}
\end{center}
\caption{}
\end{figure}
\vfill

\begin{figure}
\begin{center}
\leavevmode
\epsfxsize=14cm
\epsffile{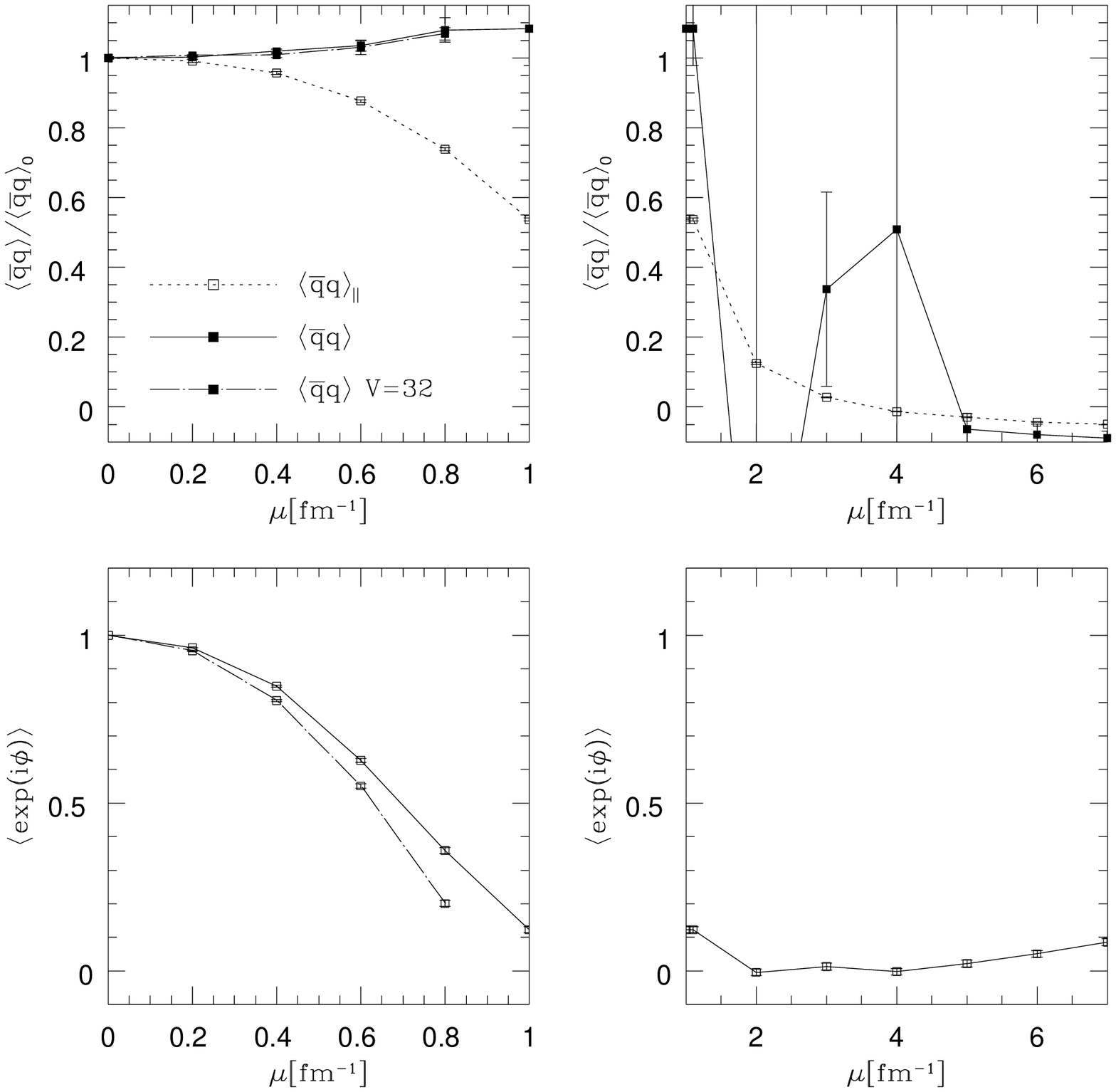}
\end{center}
\caption{}
\end{figure}
\vfill

\begin{figure}
\begin{center}
\leavevmode
\epsfxsize=14cm
\epsffile{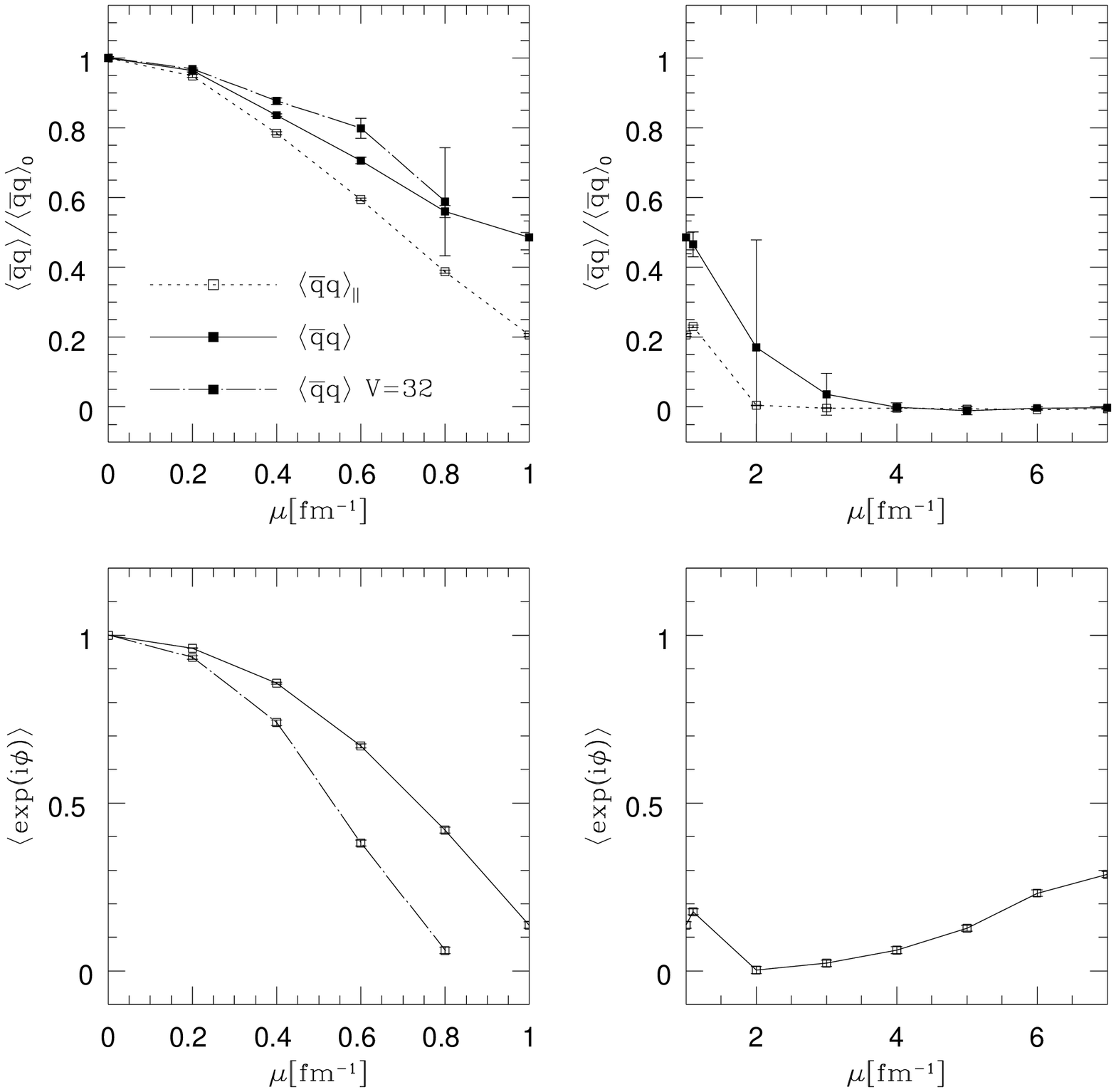}
\end{center}
\caption{}
\end{figure}
\vfill

\begin{figure}
\begin{center}
\leavevmode
\epsfxsize=8cm
\epsffile{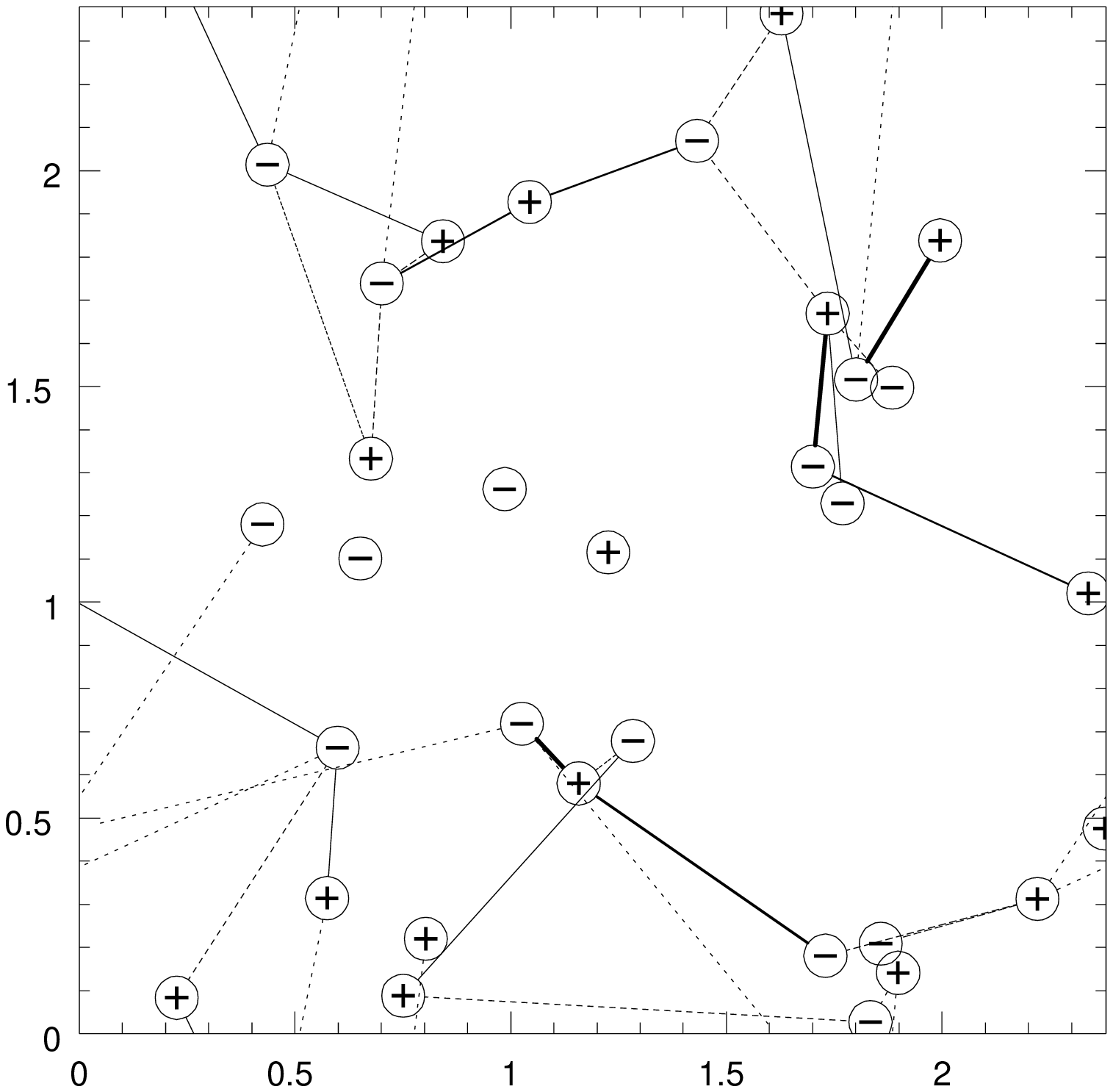}
\end{center}
\begin{center}
\leavevmode
\epsfxsize=8cm
\epsffile{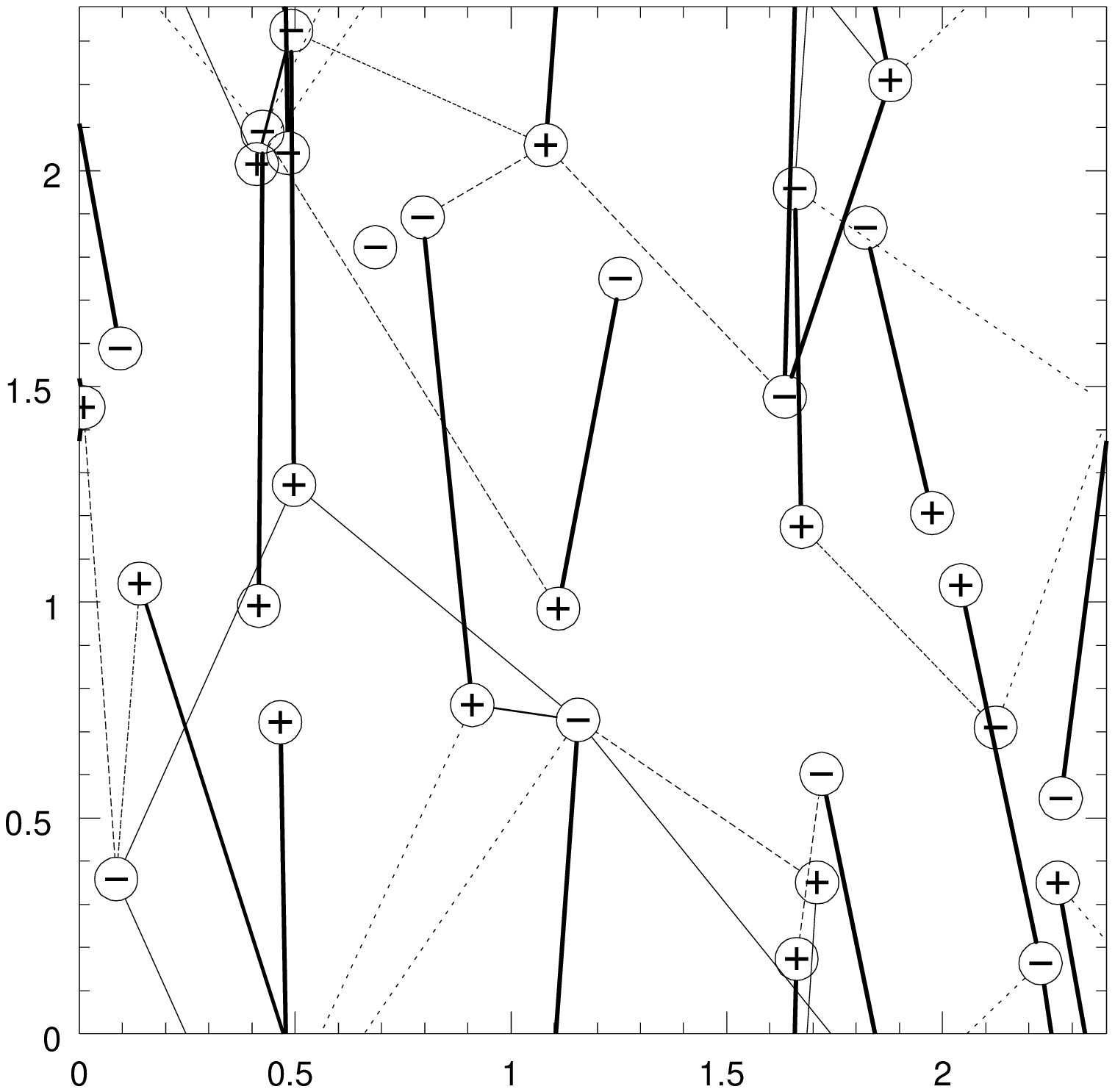}
\end{center}
\caption{}
\end{figure}
\vfill

\begin{figure}
\begin{center}
\leavevmode
\epsfxsize=14cm
\epsffile{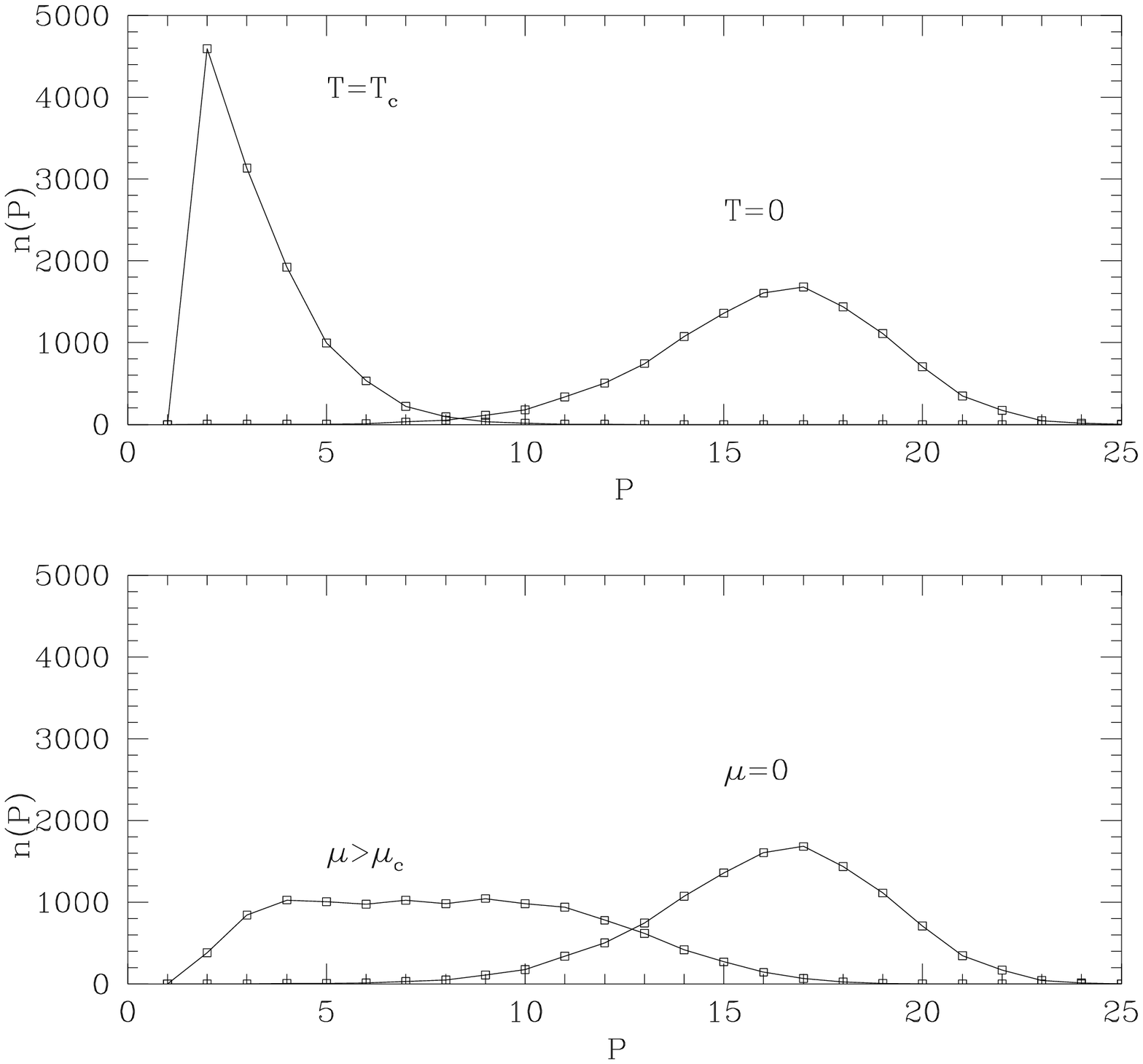}
\end{center}
\caption{}
\end{figure}
\vfill

\end{document}